\documentclass[reprint,superscriptaddress,amsmath,amssymb,aps,pre,floatfix]{revtex4-2}
\usepackage{graphicx}
\usepackage{mathtools}
\usepackage{booktabs}
\usepackage{CJKutf8}
\usepackage{bm}
\usepackage[bookmarks=false,breaklinks=true,pdfborder={0 0 0},pdfborderstyle={},backref=false,colorlinks=true]{hyperref}

\begin{document}
\title{Information-Theoretic Reliability--Safety Bounds and Finite-Alignment Dynamics in AI Knowledge Distillation}
\author{Miao-Miao Yi (\begin{CJK}{UTF8}{gbsn}易淼淼\end{CJK})}
\affiliation{Graduate School of China Academy of Engineering Physics, Beijing, 100193, China}
\author{Zhen-Yu Cai (\begin{CJK}{UTF8}{gbsn}蔡镇宇\end{CJK})}
\affiliation{Graduate School of China Academy of Engineering Physics, Beijing, 100193, China}
\author{Y -M. Du (\begin{CJK}{UTF8}{gbsn}杜亦牧\end{CJK})}
\affiliation{Graduate School of China Academy of Engineering Physics, Beijing, 100193, China}
\author{C. P. Sun (\begin{CJK}{UTF8}{gbsn}孙昌璞\end{CJK})}
\email{suncp@gscaep.ac.cn}
\affiliation{Graduate School of China Academy of Engineering Physics, Beijing, 100193, China}
\begin{abstract}
The reliability and safety of artificial intelligence (AI), together with their transfer and evolution through knowledge distillation, have attracted growing attention. We establish an information-theoretic bound on the reliability–safety trade-off and characterize the dynamics of this trade-off under knowledge distillation. We first prove a universal trade-off inequality that is independent of the specific model form. Its boundary is determined solely by a single parameter $K_h$, which quantifies the hidden layer's ability to distinguish benign from hazardous inputs. We then develop a canonical Boltzmann-machine model of AI response behavior and derive two exact relations between reliability and safety. The trade-off curve is determined by the hazard discrimination capacity $K$, while the overall answer tendency controls the operating point along the curve. These relations agree well with existing experimental data. In knowledge distillation, we find that teacher responses induce a renormalization of the student model's trade-off parameters. We further use a renormalization-group approach to study multigenerational distillation, deriving exact recursion relations and the associated renormalization-group-like flows for two representative direct intergenerational distillation protocols at arbitrary teacher–student alignment strength. These flows reveal fixed-point structures, long-term dynamics, and critical behavior, yielding testable predictions for the critical transition between capability loss and stable transmission.
\end{abstract}
\maketitle

\section{Introduction}

An ideal artificial intelligence (AI) system should possess not only the
functional reliability to provide effective responses to normal queries,
but also the safety to refuse or take appropriate safety measures when
confronted with sensitive, malicious, or hazardous queries. However,
reliability and safety impose distinct response requirements for
different types of inputs. High task reliability does not necessarily
imply high safety, suggesting that the two may be subject to intrinsic
constraints. Meanwhile, as the potential global risks associated with
the development of AI become increasingly apparent, AI safety has
received growing attention \cite{CenterForAISafety2023Statement,Bengio2024Science}. Quantitatively characterizing
the accessible range of reliability and safety, their trade-off
boundaries, and the underlying microscopic mechanisms therefore
constitutes a fundamental scientific problem for understanding the
intrinsic behavioral properties of AI models and addressing their
safety.

Meanwhile, knowledge distillation is a widely used approach for AI model
compression and capability transfer and can reduce the computational resources required by the model, in which a teacher model guides and
trains a smaller student model through its output behavior
\cite{Hinton2015arXiv,Gou2021IJCV}. Distillation transfers not only task-performing
capabilities, but also the response tendencies and behavioral boundaries
of a model across different types of inputs. Behavioral differences of
the teacher model between normal and hazardous inputs may therefore be
preserved, weakened, or altered during distillation. Thus, beyond
understanding the reliability and safety of an individual model, how
knowledge distillation modifies these properties and their mutual
constraints constitutes another fundamental problem beyond capability
transfer itself \cite{Cloud2026Nature,Alfarra2026UAI}.

This issue becomes particularly important in multigenerational
distillation. In processes such as multigenerational self-distillation
\cite{Furlanello2018ICML,Pareek2024NeurIPS}, progressive model compression \cite{Cheng2026arXiv,Son2021ICCV}, and
teacher-assistant distillation \cite{Mirzadeh2020AAAI,Son2021ICCV}, a distilled model can
further participate in training the next generation, allowing model
behaviors to be repeatedly transmitted across generations. As
distillation proceeds, the inheritance and loss of model capabilities
can accumulate from generation to generation. The two fundamental
problems above therefore naturally extend to the multigenerational
setting: How do reliability and safety evolve under multigenerational
knowledge distillation, and what factors determine their long-term
behavior?

From a statistical-physics perspective, multigenerational distillation
has a structure analogous to a renormalization process \cite{Wilson1975RMP}. Knowledge
distillation allows the outputs of a teacher model to affect the
response behavior of a student model, thereby changing its macroscopic
characteristics. When distillation is repeated across generations, the
parameters describing the macroscopic behavior of the model are
iteratively mapped, giving rise to a renormalization-group (RG)-like
parameter flow. In our previous work, we used a specific model to reveal
a trade-off relation between model reliability and safety and introduced
a teacher-student alignment interaction to describe knowledge
distillation. For a particular class of multigenerational distillation
processes in the weak-alignment regime, we studied the generational
evolution of the corresponding trade-off parameters, demonstrated that
it forms an RG-like flow, and revealed the associated fixed-point
structure and critical behavior \cite{Du2026arXiv}.

The present work extends our previous study in several directions. First, using information theory, we establish a key result: AI reliability and safety satisfy a universal trade-off inequality that is independent of the specific model form. The information-theoretic bound is determined solely by a single parameter $K_h$, which quantifies the hidden layer's ability to distinguish benign from hazardous inputs. This result applies to typical AI models  with an input–hidden-layer–response Markov structure \cite{LeCun1998ProceedingsIEEE,Krizhevsky2012NeurIPS,LeCun2015Nature}.

Second, we review and further refine a canonical statistical model of AI response behavior, which yields two exact trade-off equalities between reliability and safety \cite{Du2026arXiv}. The corresponding trade-off parameters, $K$ and $B$, are termed the hazard discrimination capacity (HDC) and the overall answer tendency (OAT), respectively. $K$ determines the trade-off curve, while $B$ controls the model's operating point along that curve. These relations agree well with existing experimental data on refusal behavior \cite{Jain2025COLM,Giorgi2026npjDigitalMedicine}. Moreover, using a specific model with hidden degrees of freedom, the present work further reveals the microscopic origins of these two trade-off
parameters and their relation to hidden-layer size.

Finally, we study
the renormalization of the student model's trade-off parameters induced by knowledge
distillation and further investigate multigenerational distillation. The present work further considers arbitrary teacher–student alignment strength and derives exact recursion dynamics for two representative direct intergenerational distillation protocols with fixed and progressive initialization, yielding the corresponding RG-like flows. By analyzing their fixed-point structures and stability, we predict the long-term evolution under different initialization protocols. For progressive initialization, we further identify the critical behavior governing the transition from capability loss to stable transmission and characterize how reliability, safety, and the trade-off parameters evolve across this transition.

This paper is organized as follows. Sec.~\ref{sec:universal_trade_off} presents one of the key results of this work, namely, that the reliability and safety of AI models satisfy a universal trade-off inequality. Sec.~\ref{sec:canonical_model} introduces a minimal statistical model of AI response behavior. Sec.~\ref{sec:exact_trade_off_canonical_model} further derives two exact reliability--safety trade-off equalities based on this model and discusses the microscopic origins of the corresponding trade-off parameters. Sec.~\ref{sec:knowledge_distillation_induced_renormalization} further investigates knowledge distillation. Secs.~\ref{sec:RG_like_flows_fixed_initialization} and \ref{RG_flow_progressive_distillation} use a renormalization-group approach to perform exact analyses of two more representative direct intergenerational distillation processes, revealing their long-term evolution and critical behavior and providing testable predictions. Finally, Sec.~\ref{sec:conclusion} summarizes the main results.

\section{A Universal Information-Theoretic Bound on the Reliability–Safety Trade-off}\label{sec:universal_trade_off}

In this section, we define the reliability and safety of AI models based on their response behavior to different types of inputs, and use information theory to establish a key result of this work: the reliability and safety of AI models satisfy a universal trade-off inequality. This inequality is independent of the specific model form and follows from the input--hidden-layer--response information-transmission structure shared by typical AI architectures. We further show that the boundary of this universal inequality is controlled by a single parameter, $K_{\bm h}$. This parameter is defined in terms of the maximum relative entropy, and quantifies the hidden layer's ability to distinguish normal from hazardous inputs.

\begin{figure}[h]
    \centering
    \includegraphics[width=\linewidth]{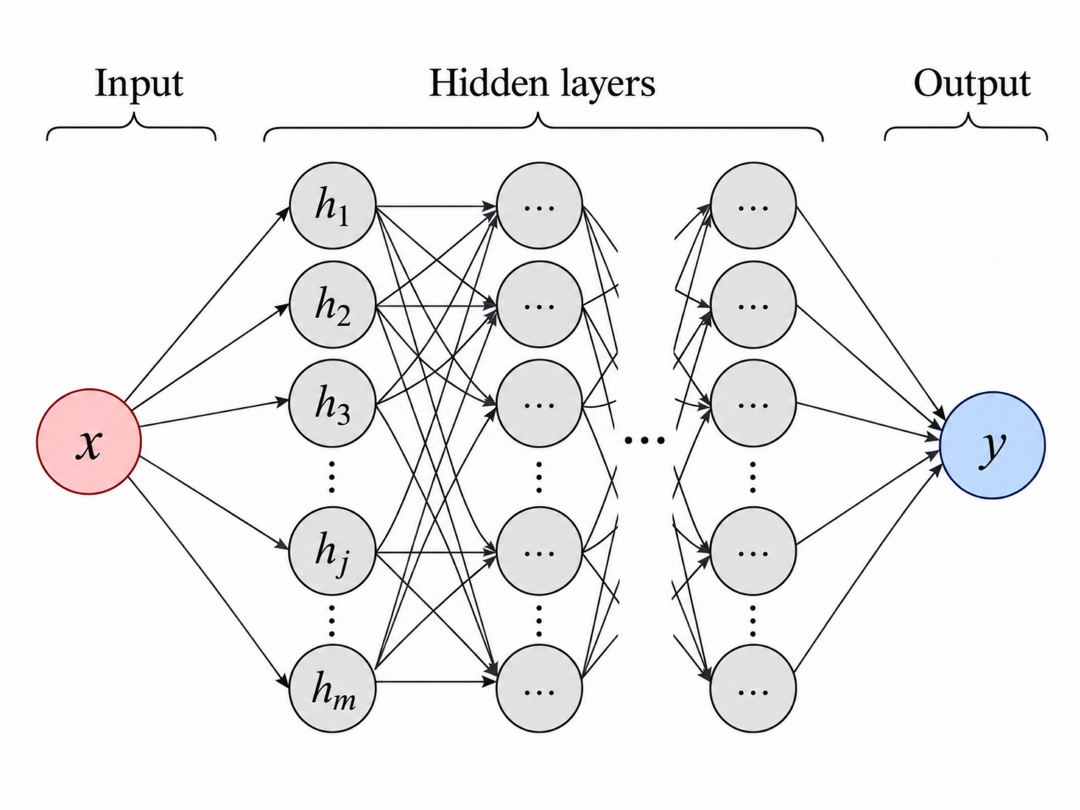}
    \caption{Basic architecture of an AI model. The input, hidden, and output layers are shown in red, gray, and blue, respectively.}
    \label{fig:1_multi_hidden_layer}
\end{figure}

In the basic architecture of typical AI models \cite{Goodfellow2016DeepLearning,LeCun2015Nature}, the
input-response process generally proceeds through an input layer, a
hidden layer, and a response layer, as shown in Fig.~\ref{fig:1_multi_hidden_layer}. To focus on
reliable and safe behavior, we coarse-grain inputs and responses into a
binary structure. The input variable $x = + 1$ denotes a benign
external input, and $x = - 1$ a hazardous input, such as a sensitive
or harmful question. The response variable $y = + 1$ denotes answering
or executing an action, and $y = - 1$ denotes refusal or entry into a
safe mode. The hidden variables are denoted as
$\bm{h} = \left( h_{1},h_{2},\cdots,h_{N} \right)^{T}$, where
$h_{j}$ is the $j$th hidden variable. Reliability and safety are
then defined by the following conditional probabilities
\begin{align}
R =&\,P\left( y = + 1 \middle| x = + 1 \right),\label{eq:reliability_definition}\\
S =&\,P\left( y = - 1 \middle| x = - 1 \right).\label{eq:safety_definition}
\end{align}
Here, the reliability $R$ is the probability that the model provides a normal response or executes an action for a normal input, whereas the safety $S$ is the probability that the model refuses to respond or blocks an action for a hazardous input.


In typical AI models, such as those based on artificial neural networks
(ANNs), input information is processed through internal hidden
representations before the final response is generated \cite{LeCun1998ProceedingsIEEE,Krizhevsky2012NeurIPS,LeCun2015Nature}.
The input $x$ does not directly determine the response layer; it first affects the hidden layer $\bm{h}$, which
then determines the response. This information flow can be represented
by the Markov chain $x \rightarrow \bm{h} \rightarrow y$
\cite{Cover2006ElementsInformationTheory}, implying
\begin{equation}\label{eq:3}
P\left( y \middle| \bm{h},x \right) = P\left( y \middle| \bm{h} \right),
\end{equation}
Thus, given the hidden-layer state $\bm{h}$, the response $y$ no
longer depends directly on the input $x$. Combining
$P\left( y \middle| x \right) = \sum_{\bm{h}}^{}{P\left( y,\bm{h} \middle| x \right)}$
with
$P\left( y,\bm{h} \middle| x \right) = P\left( y \middle| \bm{h},x \right)P\left( \bm{h} \middle| x \right)$,
we can rewrite reliability and safety as
\begin{align}
R=&\,\sum_{\bm{h}}^{}{P\left( y = + 1 \middle| \bm{h} \right)P\left(\bm{h} \middle| x = + 1 \right)},\label{eq:reliability_demonstrate}\\
S=&\,\sum_{\bm{h}}^{}{P(y = - 1|\bm{h})P\left( \bm{h} \middle| x = - 1 \right)}.\label{eq:safety_demonstrate}
\end{align}

We now show that this structure implies a universal reliability--safety
trade-off inequality. Let
$P_{\pm}\left( \bm{h} \right) \coloneqq P\left( \bm{h} \middle| x = \pm 1 \right)$
denote the probability distribution of the hidden-layer state
$\bm{h}$ for each input class. The maximum-relative entropy is
defined as
\begin{equation}
D_{\infty}\left( P_{\pm} \parallel P_{\mp} \right):=\sup_{\bm{h}}{\ln\left\lbrack \frac{P_{\pm}\left( \bm{h} \right)}{P_{\mp}\left( \bm{h} \right)} \right\rbrack}.\label{eq:D_infty_definition}
\end{equation}
Based on this, we can further obtain a universal constraint on reliability and safety that is independent of the specific form of the AI model, as stated in Theorem 1 below.

\textbf{Theorem 1.} \emph{The AI model's reliability and safety satisfy}
\begin{equation}\label{eq:unequality_trade_off}
\left| \operatorname{Logit}(R) + \operatorname{Logit}(S) \right| \leq K_{\bm h},
\end{equation}
\emph{with}
\begin{equation}\label{eq:8}
K_{\bm h}:=D_{\infty}\left( P_{+} \parallel P_{-} \right) + D_{\infty}\left( P_{-} \parallel P_{+} \right),
\end{equation}
\emph{and}
$\operatorname{Logit}(u) = \ln\left[u/(1-u)\right]$\emph{.}

\textbf{Proof.} Write $D_{+} \coloneqq D_{\infty}\left( P_{+} \parallel P_{-} \right)$,
$D_{-} \coloneqq D_{\infty}\left( P_{-} \parallel P_{+} \right)$,
$q_{\bm{h}} \coloneqq P\left( y = + 1 \middle| \bm{h} \right)$,
$p_{\bm{h}} := P\left( \bm{h} \middle| x = + 1 \right)$, and
$r_{\bm{h}} \coloneqq P\left( \bm{h} \middle| x = - 1 \right)$. By the
definition of maximum-relative entropy in Eq.~\eqref{eq:D_infty_definition}, for any
$\bm{h}$ we have
\begin{equation}\label{eq:9}
e^{- D_{-}}r_{\bm{h}} \leq p_{\bm{h}} \leq e^{D_{+}}r_{\bm{h}}.
\end{equation}
Combining Eqs.~\eqref{eq:reliability_demonstrate} and \eqref{eq:safety_demonstrate} gives
\begin{equation}
\,R=\sum_{\bm{h}}^{}{q_{\bm{h}}p_{\bm{h}}}\leq e^{D_{+}}\sum_{\bm{h}}^{}{q_{\bm{h}}r_{\bm{h}}}= e^{D_{+}}(1 - S),\label{eq:10}
\end{equation}
and
\begin{equation}
e^{- D_{-}}(1 - S)=e^{- D_{-}}\sum_{\bm{h}}^{}{q_{\bm{h}}r_{\bm{h}}}\leq\sum_{\bm{h}}^{}{q_{\bm{h}}p_{\bm{h}}}= R.\label{eq:11}
\end{equation}
Hence,
\begin{equation}\label{eq:12}
- D_{-} \leq \ln{\frac{R}{1 - S} }\leq D_{+}.
\end{equation}

Similarly, Eq.~\eqref{eq:9} gives
$\exp\left( - D_{+} \right)p_{\bm{h}} \leq r_{\bm{h}} \leq \exp{\left( D_{-} \right)p_{\bm{h}}}$.
Applying the same argument yields
\begin{equation}\label{eq:13}
- D_{+} \leq \ln\frac{S}{1 - R} \leq D_{-}.
\end{equation}

Adding Eqs.~\eqref{eq:12} and \eqref{eq:13}, we obtain
\begin{equation}\label{eq:14}
- K_{\bm{h}} \leq \operatorname{Logit}(R) + \operatorname{Logit}(S) \leq K_{\bm{h}}.
\end{equation}
which gives Eq.~\eqref{eq:unequality_trade_off}.

$\square$

Equality in Eq.~\eqref{eq:unequality_trade_off} corresponds to the limiting case in which the
response layer fully utilizes the discriminative information encoded in the hidden layer. Specifically, equality at the upper boundary
$K_{\bm h}$ requires every hidden state $\bm{h}$ producing an answer $y=+1$ to
correspond to the maximum of
$\ln\left\lbrack P_{+}\left( \bm{h} \right)/P_{-}\left( \bm{h} \right)\right\rbrack$,
and every  hidden state producing refusal responses correspond to its minimum. The lower
boundary $- K_{\bm h}$ corresponds to the opposite response mapping.

By the definition of maximum-relative entropy \cite{vanErven2014IEEETransactionsInformationTheory}, $K_{\bm h}$
characterizes the maximum discrimination between the two input classes
in the hidden layer. The hidden-layer structure therefore determines the
universal reliability--safety boundary. In particular, at a given
reliability $R$, the inequality gives the safety bound
\begin{equation}\label{eq:15}
S \leq \frac{e^{K_{\bm h}}(1 - R)}{R + e^{K_{\bm h}}(1 - R)}.
\end{equation}
For a fixed hidden layer, every possible reliability and safety
$(R,S)$ is constrained by the boundary curve in Eq.~\eqref{eq:unequality_trade_off}, as
illustrated in Fig.~\ref{fig:2_trade_off_universal_and_exact} (b). The maximum attainable region can expand only if
the hidden layer better distinguishes normal inputs and hazardous
inputs. This provides a general reliability--safety constraint
independent of the specific AI model.

Similar trade-off bound occur widely in physical systems. For example, power
and efficiency in finite-time thermodynamics are constrained by a
power-efficiency boundary \cite{Shiraishi2016PRL,Pietzonka2018PRL,ma1,ma2,ma3,ma4}. Under fixed system conditions,
different operating protocols select different points within the
attainable region, with optimal protocols reaching its boundary. Here,
$K_{\bm h}$ imposes an analogous constraint: it is determined by
hidden-layer discrimination and sets an information-theoretic constraint on jointly achievable reliability and safety.

\section{A Canonical Boltzmann-Machine Model for Input-Response behavior}\label{sec:canonical_model}
To reveal further properties of the relation between reliability and
safety, and study its evolution during knowledge distillation in an
exactly solvable setting, we review a canonical Boltzmann-machine model of AI input–response behavior introduced in our previous work \cite{Du2026arXiv}. Here, we present it in a more intuitive and compact form. Similar canonical descriptions
are widely used in Boltzmann machines, Markov/conditional random fields,
and energy-based neural models \cite{Hinton1986PDP,LeCun2007PredictingStructuredData,Ackley1985CognitiveScience,Lafferty2001ICML,Torlai2016PRB}.

Specifically, the joint distribution of the input $x$, hidden states
$\bm{h}$, and response $y$ is taken to have the canonical form
\begin{equation}\label{eq:16}
P\left( y,\bm{h},x \right) = \frac{1}{Z_{0}}\exp\left\lbrack - \beta\widetilde{E}\left( y,\bm{h},x \right) \right\rbrack,
\end{equation}
where $\widetilde{E}\left( y,\bm{h,}x \right)$ is the energy of
the system state $y$, $\bm{h}$ for input $x$, $\beta$ is the
inverse temperature characterizing noise strength, and $Z_{0}$ is the
partition function.

Since we focus on the final answer or refusal rather than the specific
hidden degrees of freedom $\bm{h}$, we sum over $\bm{h}$ to
obtain the reduced input-response distribution
\begin{equation}
P(y,x)=\sum_{\bm{h}}^{}{P\left( y,\bm{h},x \right)} = \frac{1}{Z}\exp\left\lbrack - \beta E(y,x) \right\rbrack,\label{eq:canonical_form_sum_over_h}
\end{equation}
where
\begin{equation}\label{eq:energy_sum_over_h}
E(y,x) = -\frac{1}{\beta}\ln{\sum_{h}^{}{\exp\left\lbrack - \beta \widetilde{E}\left( y,\bm{h},x \right) \right\rbrack}}.
\end{equation}
For binary input and response variables $x,y = \pm 1$, any energy
function $E(y,x)$ can be written as
\begin{equation}\label{eq:reduced_energy_abJ_express}
E(y,x) = - ax - by - Jxy,
\end{equation}
where the state-independent constant has been omitted. The corresponding
partition function is
\begin{equation}\label{eq:partion_function}
Z = 2\left\lbrack e^{\beta a}\cosh{\beta(b + J)} + e^{- \beta a}\cosh{\beta(b - J)}\right\rbrack.
\end{equation}
Here, the parameter $b$ characterizes the overall answer tendency of the model. A positive $b$ makes the system more likely to produce a normal response, whereas the opposite holds for $b < 0$. The parameter $J$ characterizes the input-response matching strength. 
For $J > 0$, increasing $J$ makes the model more likely to respond normally to normal inputs and to block hazardous inputs.

The parameter $a$ characterizes the overall input tendency of the model. In an AI model,
inputs are typically specified by external users. Therefore the input probability
$P(x)$ should be externally determined rather than affected by the
internal parameters $b$ and $J$. Let $\rho$ be the externally
specified probability of a hazardous input. To ensure that $P(x)$ in the above statistical model is consistent with this externally specified probability, i.e.,
$P(x = - 1) = \rho$, the parameter $a$ must satisfy
\begin{equation}
a=\frac{1}{2\beta}\ln\frac{1 - \rho}{\rho}-\frac{1}{2\beta}\ln\frac{\cosh\left\lbrack \beta(b + J) \right\rbrack}{\cosh\left\lbrack \beta(b - J) \right\rbrack}.\label{eq:21}
\end{equation}
Thus, $a$ is determined self-consistently by the external input
probability $\rho$. For $a > 0$, increasing $a$ favors benign
inputs $x = + 1$; $a < 0$ favors the opposite.

\section{Exact reliability--safety Trade-off and microscopic origins}\label{sec:exact_trade_off_canonical_model}

In this section, we discuss the reliability and safety of AI models
based on the canonical statistical model introduced above. Beyond the
general constraint in the trade-off inequality \eqref{eq:unequality_trade_off}, the canonical
model yields two exact trade-off equalities, controlled by the
macroscopic parameter $K = 4\beta J$ and $B = 4\beta b$. We call them the hazard discrimination capacity (HDC) and the overall answer tendency (OAT), respectively. Similar equalities were first obtained in our previous work \cite{Du2026arXiv}. The reliability--safety trade-off curve is determined solely by the HDC $K$, while the OAT $B$ controls the model's position along this curve. This result is consistent with existing experimental data on refusal
behavior in current AI models \cite{Jain2025COLM,Giorgi2026npjDigitalMedicine}. Furthermore, by
considering an explicit model with hidden degrees of freedom, this paper further reveals
the microscopic origins of the two trade-off parameters and their
relation to the hidden-layer size.

The canonical model introduced in Sec.~\ref{sec:canonical_model} gives the response probability for a specified input
$x$,
\begin{align}
P\left(y\middle| x \right)=\frac{\exp\left\lbrack \beta y(b + Jx) \right\rbrack}{2\operatorname{cosh}\left\lbrack\beta(b+Jx) \right\rbrack}.\label{eq:22}
\end{align}
Using the definitions of reliability and safety in Eqs.~\eqref{eq:reliability_definition} and
\eqref{eq:safety_definition}, we obtain
\begin{align}
    R=&\,\frac{1}{1 + \exp\left\lbrack - 2\beta(b + J) \right\rbrack},\label{eq:reliability_canonical_model}\\
    S=&\,\frac{1}{1 +\exp\left\lbrack - 2\beta(J - b) \right\rbrack}.\label{eq:safety_canonical_model}
\end{align}
Both $R$ and $S$ are independent of the input-bias parameter $a$.
Reliability and safety are therefore determined solely by internal model
parameters, they describe the model's ability to handle
benign and hazardous inputs, independently of the external input
distribution.

\begin{figure}[tb]
    \centering
    \includegraphics[width=0.7\linewidth]{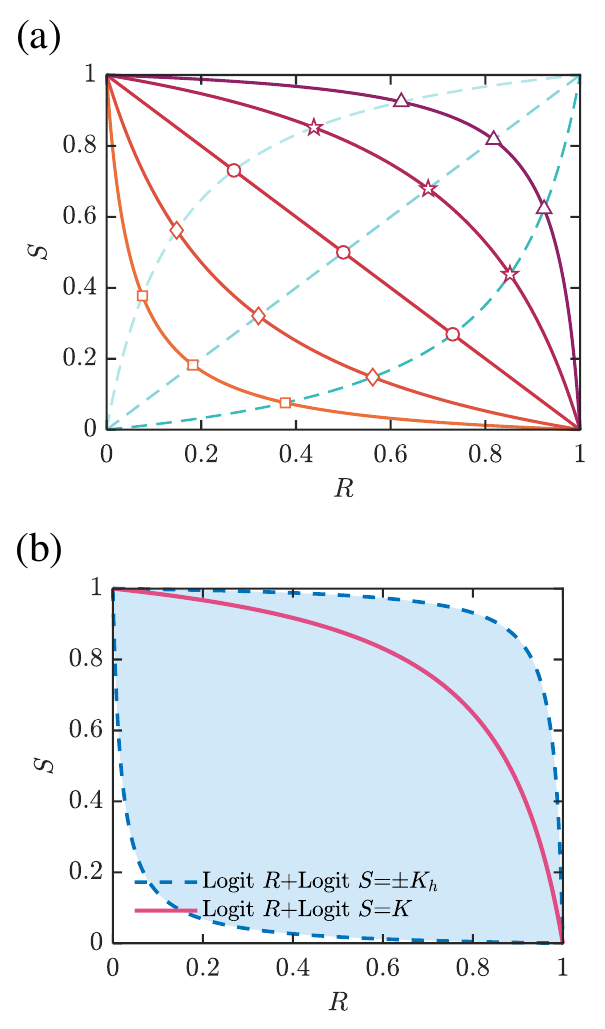}
    \caption{(a) Exact trade-off relations in Eqs.~\eqref{eq:equality_trade_off_K_canonical} and \eqref{eq:equality_trade_off_B_canonical}. Solid curves represent Eq.~\eqref{eq:equality_trade_off_K_canonical}. Different colors denote different $K$ values; from orange to purple, $K = - 3,\ - 1.5,\ 0,\ 1.5,\ 3$. Dashed curves represent Eq.~\eqref{eq:equality_trade_off_B_canonical}. Their colors denote different $B$ values; from light to dark, $B = - 2,\ 0,\ 2$. (b) The region permitted by the information-theoretic bound and trade-off curve. Blue dashed curves show the universal boundary determined by $K_{\bm h}$; the blue shading denotes the region permitted by the bound. For a model with a given hidden-layer structure, reliability and safety are restricted to a trade-off curve determined by $K$.}
    \label{fig:2_trade_off_universal_and_exact}
\end{figure}

From Eqs.~\eqref{eq:reliability_canonical_model} and \eqref{eq:safety_canonical_model}, reliability $R$ and safety $S$
satisfy two exact trade-off equalities
\begin{equation}
    \operatorname{Logit}(R)+\operatorname{Logit}(S) =K,\label{eq:equality_trade_off_K_canonical}
\end{equation}
and
\begin{equation}
    \operatorname{Logit}(R)-\operatorname{Logit}(S) =B,\label{eq:equality_trade_off_B_canonical}
\end{equation}
where
\begin{equation}\label{eq:27}
K := 4\beta J,\qquad B := 4\beta b.
\end{equation}
Associated with the input-response coupling $J$, $K$ characterizes
the AI model's ability to adjust its response behavior
according to the input class, we therefore call $K$ the hazard
discrimination capacity (HDC). Similarly, we refer to $B$, associated
with $b$, as the overall answer tendency (OAT). These parameters play
distinct physical roles, $K$ determines the reliability--safety
trade-off boundary, while $B$ selects the balance between reliability
and safety along a fixed boundary, as shown in Fig.~\ref{fig:2_trade_off_universal_and_exact} (a). Increasing $K$
expands the attainable region, allowing reliability and safety to
improve together, whereas increasing $B$ improves reliability while
reducing safety.


Notably, the exact equality \eqref{eq:equality_trade_off_K_canonical} is directly related to the universal
inequality \eqref{eq:unequality_trade_off} in Sec.~\ref{sec:universal_trade_off}; together they imply
$|K| \leq K_{\bm{h}}$. As shown in Fig.~\ref{fig:2_trade_off_universal_and_exact} (b), $K_{\bm h}$ sets a
universal reliability--safety boundary. Equation \eqref{eq:equality_trade_off_K_canonical} shows that,
$R$ and $S$ cannot vary arbitrarily within this region but lie on an
exact trade-off curve determined by $K$. When $|K| = K_{\bm h}$, this
curve reaches the universal boundary. The exact relation at fixed $K$
is first obtained in our earlier work \cite{Du2026arXiv}. In the present
work, Eq.~\eqref{eq:unequality_trade_off} extends this result to a universal constraint that is
independent of the particular statistical model.

\begin{figure}
    \centering
    \includegraphics[width=0.85\linewidth]{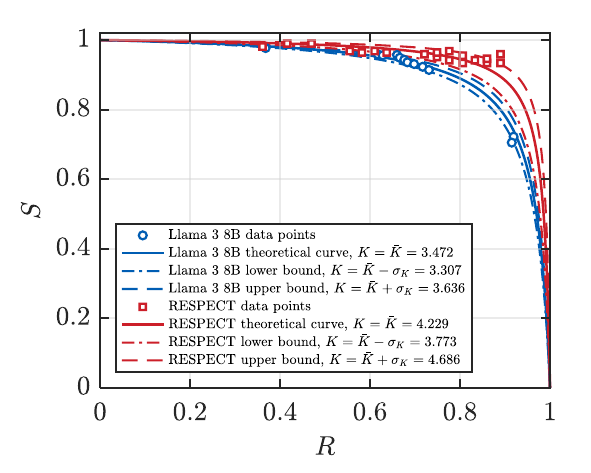}
    \caption{Experimental validation of the reliability--safety trade-off curves. Blue circles and red squares are data digitized from Fig.~2 of Ref.~\cite{Jain2025COLM} and Fig.~1 of Ref.~\cite{Giorgi2026npjDigitalMedicine}, respectively. Solid curves of the corresponding colors use the mean HDC $\overline{K}$; dashed curves use $K \pm \sigma_{K}$, where $\sigma_{K}$ is the standard deviation calculated from the digitized data. Detailed discussions and digitized data are provided in Appendix~\ref{app:digitized_data}.}
    \label{fig:3_data_trade_off}
\end{figure}

These exact trade-off relations are consistent with refusal data from
existing large language models. According to Eqs.~\eqref{eq:equality_trade_off_K_canonical} and \eqref{eq:equality_trade_off_B_canonical},
varying the OAT $B$ trades reliability against safety, while the two
remain on a trade-off curve determined by a fixed $K$. In the
refusal-token experiment of Ref.~\cite{Jain2025COLM}, the overall answer
probability of the same model was continuously varied by changing the
answer-refusal threshold. Similarly, the recent RESPECT experiment
varied the refusal threshold and obtained a Refusal-Utility Curve
relating the utility of answering appropriate questions to the
appropriate-refusal rate for inappropriate questions
\cite{Giorgi2026npjDigitalMedicine}. Mapping these two data sets onto the
reliability $R$ and safety $S$ defined here, we find that the data
at different thresholds approximately follow curves with fixed $K$, as
shown in Fig.~\ref{fig:3_data_trade_off}. Other experiments on safety alignment, overrefusal, and
safety-utility trade-offs exhibit similar compromises between
reliability and safety \cite{Cui2025ICML,Wolf2025ICLRWorkshop}.

Next, we use an explicit model with hidden degrees of freedom to relate
$J$ and $b$ to the hidden layer and reveal the microscopic origins
of $K$ and $B$. We show that, both parameters depend on the
hidden-layer size $N$, with the detailed dependence determined jointly
by interlayer couplings and interactions within the hidden layer.

Specifically, we consider a Boltzmann machine with $N$ hidden degrees
of freedom and energy
\begin{align}
\widetilde{E}\left( y,\bm{h},x \right) =&\,- a_0x - \sum_{j}^{N}\left( U_{j}xh_{j} + V_{j}yh_{j} + W_{j}h_{j} \right) \notag\\
&\,-\frac{1}{2}\sum_{ij}^{N}{Q_{ij}h_{i}h_{j}}.\label{eq:28}
\end{align}
Here, we set $h_{j} = \pm 1$. The couplings $U_{j}$ and $V_{j}$
connect the input $x$ and output $y$, respectively, to the hidden
degree of freedom $h_{j}$. The parameter $W_{j}$ biases the hidden
state toward $h_{j} = + 1$, while $Q_{ij} = Q_{ji}$ (we consider $Q_{ii}\equiv0$) is the coupling
between hidden degrees of freedom $h_{i}$ and $h_{j}$. The number
$N$ denotes the number of hidden degrees of freedom and characterizes the size of the hidden layer.

Summing over the hidden degrees of freedom yields the reduced canonical
distribution involving only $x$ and $y$. The
parameters $J$ and $b$ in the reduced energy in Eq.~\eqref{eq:reduced_energy_abJ_express} are jointly determined by the hidden-layer parameters $U_{j}$, $V_{j}$, $W_{j}$, and $Q_{ij}$; the explicit
relations are given in Appendix~\ref{sec:microscopic_mechanism_Appendix}. To further illustrate their physical origins, we
consider the weak-coupling regime
$\beta U_{j},\beta V_{j},\beta W_{j},\beta NQ_{ij} \ll 1$, the corresponding relations can then be written as
\begin{equation}\label{eq:b_microscopic_expression_main_text}
b \simeq \beta N\left\langle VW \right\rangle + \beta^{2}N^{2}\left\langle VQW \right\rangle,
\end{equation}
and
\begin{equation}\label{eq:J_microscopic_expression_main_text}
J \simeq \beta N\left\langle UV \right\rangle + \beta^{2}N^{2}\left\langle UQV \right\rangle.
\end{equation}
Here, $\langle \cdot \rangle$ denotes averages over hidden-layer
parameters, such as
$\left\langle VW \right\rangle = \sum_{j}^{}{V_{j}W_{j}}/N$
and
$\left\langle VQW \right\rangle = \sum_{ij}^{}{V_{i}Q_{ij}W_{j}}/N^{2}$.
The remaining averages are defined analogously in Appendix~\ref{sec:microscopic_mechanism_Appendix}.

Equations \eqref{eq:b_microscopic_expression_main_text} and \eqref{eq:J_microscopic_expression_main_text} show that these terms have different
origins. The product $U_{j}V_{j}$ describes transmission of input
information through the hidden degrees of freedom to the response layer.
Its accumulated contribution generates the input-response matching
strength $J$, and hence the HDC $K$. The product $V_{j}W_{j}$
transmits the hidden-layer bias to the output, generating the overall
answer tendency $b$ and the corresponding $B$. Hidden interactions
$Q_{ij}$ add cooperative contributions between hidden degrees of
freedom.

\begin{figure}
    \centering
    \includegraphics[width=0.65\linewidth]{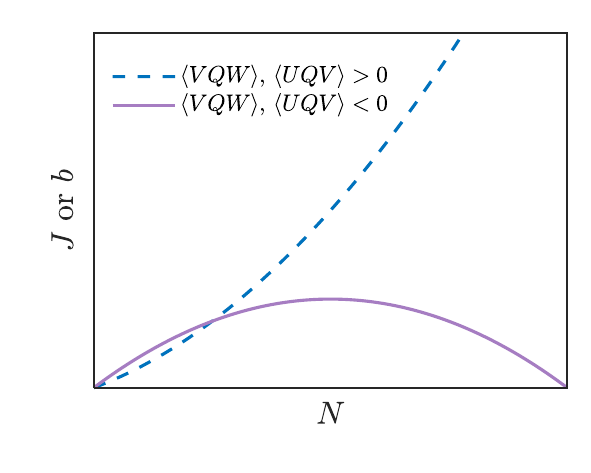}
    \caption{Schematic size dependence of $J$ or $b$ on $N$ for different accumulated hidden-layer parameters $\left\langle VQW \right\rangle,\langle UQV\rangle$, with $\left\langle VW \right\rangle,\left\langle UV \right\rangle > 0$.}
    \label{fig:4_size_dependence}
\end{figure}

When
$\left\langle VW \right\rangle,\ \left\langle UV \right\rangle,\ \left\langle VQW \right\rangle,\langle UQV\rangle > 0$,
the parameters $J$ and $b$, and hence the corresponding $K$ and $B$, increase with
hidden-layer size $N$. If the interlayer couplings $U_{j}$ and
$V_{j}$ dominate, the leading dependence scales with $N$.
Appreciable contributions from hidden-layer interactions could produce
higher-order size dependence, such as $N^{2}$ scaling. Conversely,
when
$\left\langle VQW \right\rangle,\left\langle UQV \right\rangle < 0$,
higher-order terms may suppress the growth of $J$ and $b$ with model size $N$, as illustrated in Fig.~\ref{fig:4_size_dependence}.

\section{Renormalization of the reliability--safety Trade-off During Knowledge Distillation}\label{sec:knowledge_distillation_induced_renormalization}


In this section, based on the model introduced in Sec.~\ref{sec:canonical_model}, we further
discuss knowledge distillation, in which a larger teacher model trains a
smaller student model through its output information. We show that the
teacher's responses modify the student's
energy function through distillation and thereby induce a
renormalization of its reliability--safety trade-off parameters.

\begin{figure}[tb]
    \centering
    \includegraphics[width=\linewidth]{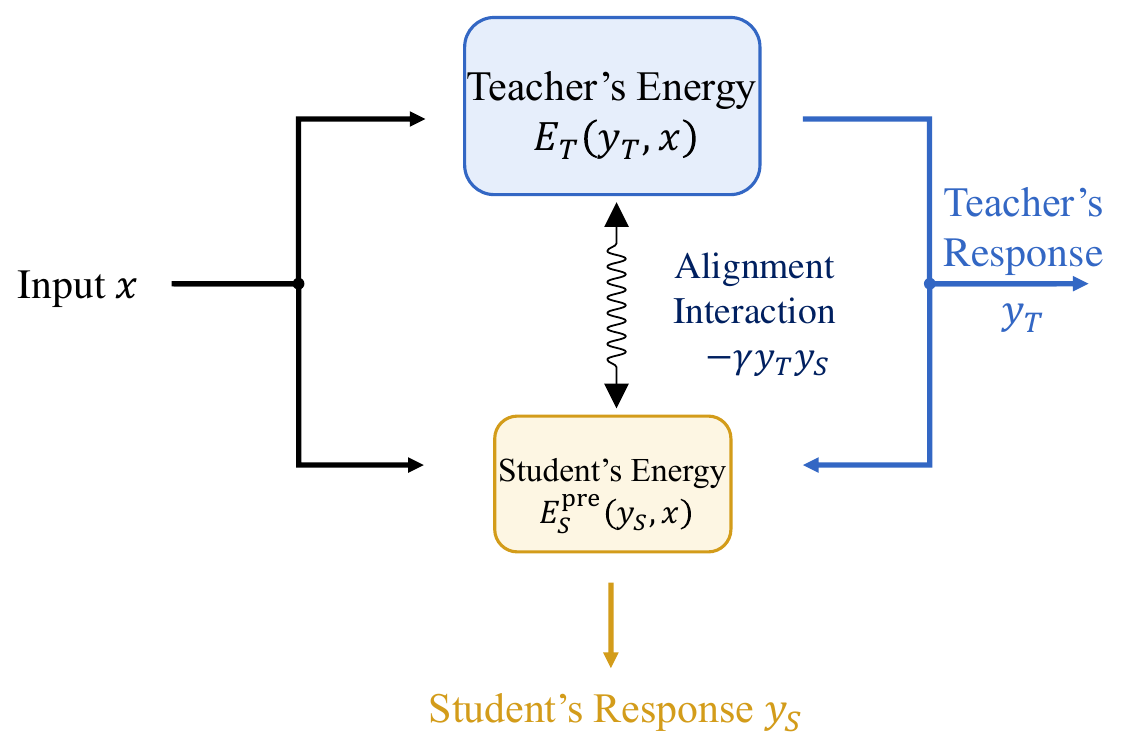}
    \caption{Joint canonical statistical model describing knowledge distillation. A large teacher model trains a smaller student through its outputs.}
    \label{fig: 5_knowledge_distillation}
\end{figure}

As illustrated in Fig.~\ref{fig: 5_knowledge_distillation}, Knowledge distillation involves two models: a
teacher and a student. Let binary variables $y_{T}$ and $y_{S}$
denote their respective answer or refusal responses to an external input
$x$. The teacher's energy is written as
$E_{T}\left( y_{T},x \right) = - a_{T}x - b_{T}y_{T} - J_{T}xy_{T}$,
and the student's pre-distillation energy is
$E_{S}^{\mathrm{pre}}\left( y_{S},x \right) = - a_{S}x - b_{S}y_{S} - J_{S}xy_{S}$.
The conditional response probabilities of both models have the form of
Eq.~\eqref{eq:22}.

During Knowledge distillation, the student's response
depends not only on the external input but also on training information
provided by the teacher's response. To describe the
distillation of the student by the teacher, we introduce a
teacher--student alignment interaction
$\beta V_{T \leftrightarrow S} = - \gamma y_{T}y_{S}$. For a given input
$x$ and teacher response $y_{T}$, the student's
conditional response probability is therefore
\begin{equation}\label{eq:31}
P_{S}\left( y_{S} \middle| x,y_{T} \right) = \frac{\exp\left\lbrack - \beta E_{S}^{\mathrm{pre}}\left( y_{S},x \right) + \gamma y_{T}y_{S} \right\rbrack}{\sum_{y_{S}'}^{}{\exp\left\lbrack - \beta E_{S}^{\mathrm{pre}}\left( y_{S}',x \right) + \gamma y_{T}y_{S}' \right\rbrack}}.
\end{equation}
Here, $\gamma$ denotes the alignment strength and characterizes the
strength of the teacher's constraint on the student. As
$\gamma$ increases, the student becomes more likely to produce
responses consistent with those of the teacher. This alignment
interaction therefore provides a coarse-grained description of the
teacher--student behavioral consistency imposed during knowledge
distillation. It corresponds to existing distillation mechanisms in
which knowledge or behavior is transferred by matching output
distributions \cite{Hinton2015arXiv}, reasoning processes \cite{Hsieh2023ACLFindings}, or behavioral responses \cite{Cloud2026Nature}.

After distillation, the teacher response $y_{T}$, which acts as an
intermediate degree of freedom during training, is averaged out. The
student's final response distribution is then
\begin{equation}\label{eq:post_probability_definition}
P_{S}^{\mathrm{post}}\left( y_{S} \middle| x \right) = \sum_{y_{T}}^{}{P_{S}\left( y_{S} \middle| x,y_{T} \right)P_{T}\left( y_{T} \middle| x \right)},
\end{equation}
this distribution retains the canonical form of Eq.~\eqref{eq:canonical_form_sum_over_h},
\begin{equation}\label{eq:retain_canonical_form}
P_{S}^{\mathrm{post}}\left( y_{S} \middle| x \right) = \frac{\exp\left\lbrack - \beta E_{S}^{\mathrm{post}}\left( y_{S},x \right) \right\rbrack}{\sum_{y_S'}^{}{\exp\left\lbrack - \beta E_{S}^{\mathrm{post}}\left( y_{S}',x \right) \right\rbrack}}.
\end{equation}
where
\begin{equation}\label{eq:reduced_post_distillation_energy}
E_{S}^{\mathrm{post}}\left( y_{S},x \right) := - a_{S}^{\mathrm{post}}x - b_{S}^{\mathrm{post}}y_{S} - J_{S}^{\mathrm{post}}xy_{S}.
\end{equation}
The teacher's response information is transferred
to the student through the alignment interaction, renormalizing the
student's pre-distillation energy parameters
$\left( b_{S}^{\mathrm{pre}},J_{S}^{\mathrm{pre}} \right)$ to
$\left( b_{S}^{\mathrm{post}},J_{S}^{\mathrm{post}} \right)$.

Because the post-distillation response distribution in Eq.~\eqref{eq:retain_canonical_form} retains the
same canonical form, its reliability $R_{S}^{\mathrm{post}}$ and safety
$S_{S}^{\mathrm{post}}$ still satisfy the two exact trade-off equalities of
Sec.~\ref{sec:exact_trade_off_canonical_model}. We now give the post-distillation trade-off parameters. Define
the trade-off parameters of the teacher, the pre-distillation student,
and the post-distillation student as
$\left( B_{T},K_{T} \right):=4\beta\left( b_{T},J_{T} \right)$,
$\left( B_{S}^{\mathrm{pre}},K_{S}^{\mathrm{pre}} \right):=4\beta\left( b_{S}^{\mathrm{pre}},J_{S}^{\mathrm{pre}} \right)$,
and
$\left( B_{S}^{\mathrm{post}},K_{S}^{\mathrm{post}} \right):=4\beta\left( b_{S}^{\mathrm{post}},J_{S}^{\mathrm{post}} \right)$,
respectively. By evaluating the parameters in Eq.~\eqref{eq:reduced_post_distillation_energy}, we have
\begin{equation}\label{eq:35}
B_{S}^{\mathrm{post}} = 2\left\lbrack {\operatorname{arctanh}}M^{+} + \operatorname{arctanh}M^{-} \right\rbrack,
\end{equation}
and
\begin{equation}\label{eq:36}
K_{S}^{\mathrm{post}} = 2\left\lbrack \operatorname{arctanh}M^{+} - \operatorname{arctanh}M^{-} \right\rbrack.
\end{equation}
Here,
\begin{align}
M^{\pm} &:=\frac{1 + \tanh H_{T}^{\pm}}{2}\tanh\left( H_{S}^{pre, \pm} + \gamma \right) \notag\\
&\quad + \frac{1 - \tanh H_{T}^{\pm}}{2}\tanh\left( H_{S}^{pre, \pm} - \gamma \right),\label{eq:37}
\end{align}
with $H_{T}^{\pm}:=\left( B_{T} \pm K_{T} \right)/4$ and
$H_{S}^{\mathrm{pre}, \pm}:=\left( B_{S}^{\mathrm{pre}} \pm K_{S}^{\mathrm{pre}} \right)/4$.

Equations~\eqref{eq:35}--\eqref{eq:37} show that the post-distillation parameters are jointly determined
by the teacher parameters $\left( B_{T},K_{T} \right)$,
pre-distillation student parameters
$\left( B_{S}^{\mathrm{pre}},K_{S}^{\mathrm{pre}} \right)$, and alignment strength
$\gamma$. We therefore write the single-step distillation process in
Eqs.~\eqref{eq:35}-\eqref{eq:37} as the following map of the trade-off parameters
\begin{equation}\label{eq:RG_flow_single_round}
\left( B_{S}^{\mathrm{post}},K_{S}^{\mathrm{post}} \right) = \mathcal{R}_{\gamma}\left( B_{T},K_{T},B_{S}^{\mathrm{pre}},K_{S}^{\mathrm{pre}} \right),
\end{equation}
This map shows that knowledge distillation transfers the
teacher's response information to the student through
teacher--student alignment and thereby renormalizes the
student's OAT $B$ and HDC $K$ according to
$\left( B_{S}^{\mathrm{pre}},K_{S}^{\mathrm{pre}} \right) \rightarrow \left( B_{S}^{\mathrm{post}},K_{S}^{\mathrm{post}} \right)$.

\section{Renormalization-group-like Flows in Multigenerational Distillation}\label{sec:RG_like_flows_fixed_initialization}

In this section and the next, we further consider a multigenerational
distillation process in which the model obtained after one generation of
distillation becomes the teacher for the next generation. Because each
student is trained directly by the model from the preceding generation,
we refer to this process as \emph{direct intergenerational
distillation}. Similar multigenerational protocols have appeared in
related approaches such as Born-Again Networks and Progressive
Distillation \cite{Furlanello2018ICML,Cheng2026arXiv}. Let $K_{n}^{\mathrm{post}}$ and $B_{n}^{\mathrm{post}}$
denote the post-distillation HDC and OAT of generation $n$, and let
$K_{n}^{\mathrm{pre}}$ and $B_{n}^{\mathrm{pre}}$ denote the corresponding
pre-distillation trade-off parameters. According to Eq.~\eqref{eq:RG_flow_single_round}, direct
intergenerational distillation obeys the following RG-like flow
\begin{equation}\label{eq:39}
\left( B_{n}^{\mathrm{post}},K_{n}^{\mathrm{post}} \right) = \mathcal{R}_{\gamma}\left( B_{n - 1}^{\mathrm{post}},K_{n - 1}^{\mathrm{post}},B_{n}^{\mathrm{pre}},K_{n}^{\mathrm{pre}} \right).
\end{equation}
This equation describes the generation-by-generation evolution of the
reliability--safety trade-off parameters.

Different alignment strengths and different pre-distillation student
initialization parameters $B_{n}^{\mathrm{pre}}$, $K_{n}^{\mathrm{pre}}$ can drive the
trade-off parameters toward different long-term states, corresponding to
capability decay, stable transmission, or further enhancement. We
therefore analyze the fixed points of the RG-like flow and their
stability to predict the long-term behavior of multigenerational
distillation. Compared with our earlier work \cite{Du2026arXiv}, here and in
the next section we consider a broader range of representative student
initializations and solve the corresponding RG-like flows exactly at
arbitrary alignment strength. Specifically, we consider fixed
initialization in this section and progressive initialization in the
next, and examine the evolution of reliability, safety, and the
associated trade-off parameters across generations.

We first consider the case in which the pre-distillation student
parameters remain fixed, $J_{n}^{\mathrm{pre}} = J^{\mathrm{pre}}$ and
$b_{n}^{\mathrm{pre}} = b^{\mathrm{pre}}$, independent of generation. This situation
represents multigenerational distillation in which the same
student-model configuration and initialization conditions are used
before distillation at every generation
\cite{Furlanello2018ICML,Pareek2024NeurIPS}.

\begin{figure}
    \centering
    \includegraphics[width=1\linewidth]{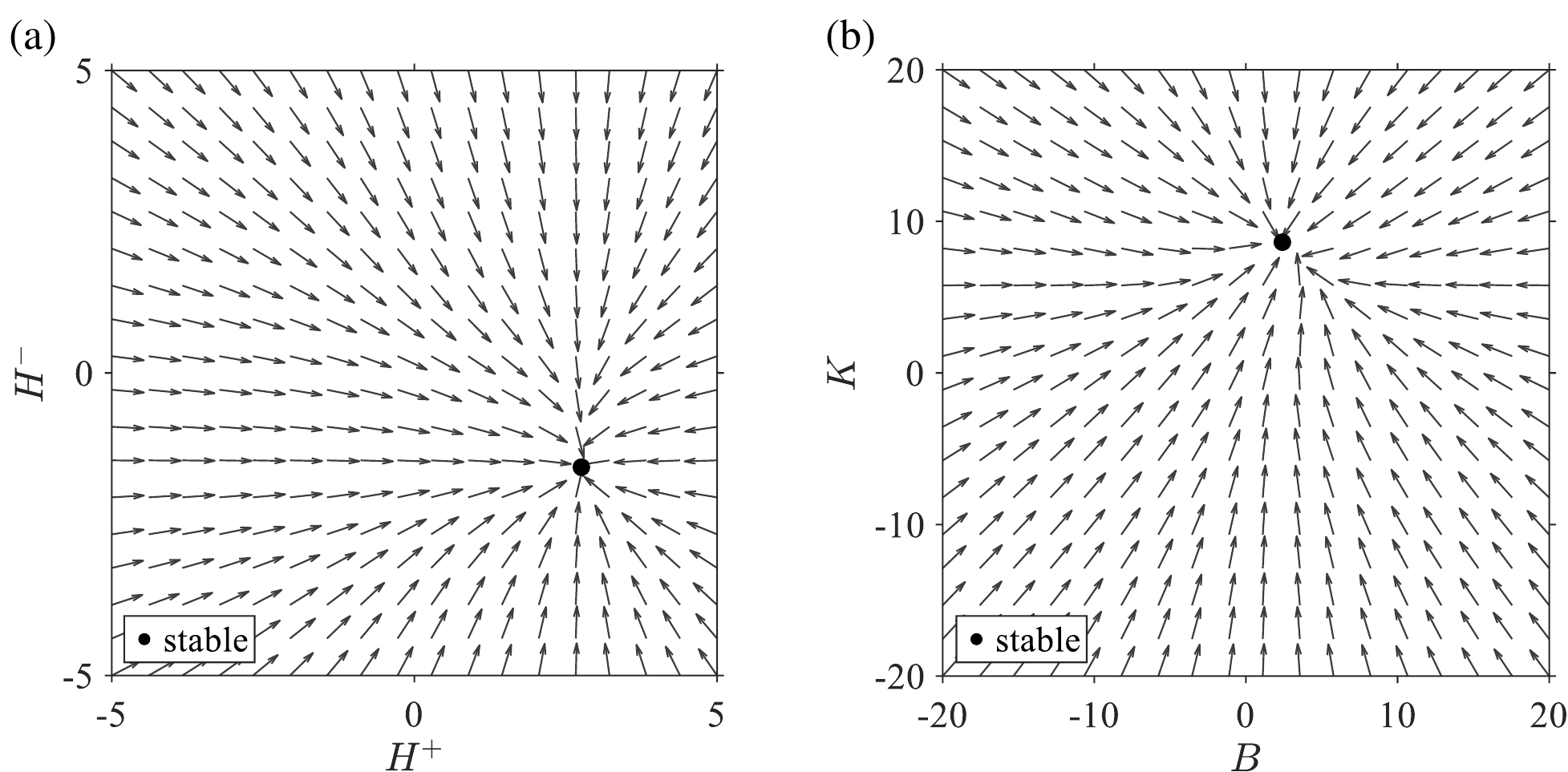}
    \caption{Fixed point and RG-like flow for $\gamma = 0.8,\ h^{+} = 2,\ h^{-} = - 1$. The unique stable fixed point is marked by a filled black circle.}
    \label{fig: 6_RG_flow_fixed_initialization}
\end{figure}

Introducing
$H_{n}^{\pm} = \left( B_{n}^{\mathrm{post}} \pm K_{n}^{\mathrm{post}} \right)/4$
and $h^{\pm} = \beta\left( b^{\mathrm{pre}} \pm J^{\mathrm{pre}} \right)$, Eq.~\eqref{eq:39}
reduces to two independent one-dimensional RG-like flows,
\begin{align}\label{eq:RG_flow_direct_distillation}
H_{n}^{\pm}=&\,\operatorname{arctanh}\Bigg\lbrack \frac{1+\tanh H_{n - 1}^{\pm}}{2}\tanh\left( h^{\pm}+\gamma \right)\notag\\
&\,+\frac{1 - \tanh H_{n - 1}^{\pm}}{2}\tanh\left( h^{\pm} - \gamma \right) \Bigg\rbrack.
\end{align}
Each flow has a unique stable fixed point,
\begin{equation}\label{eq:reliability_demonstrate1}
H_{*}^{\pm} = h^{\pm} + \operatorname{arctanh}\left\lbrack \tanh h^{\pm}\tanh\gamma \right\rbrack.
\end{equation}
The corresponding fixed point of trade-off parameters $\left( B_{*},K_{*} \right)$
is therefore given by
\begin{align}
B_{*} &= 2\left( H_{*}^{+} + H_{*}^{-} \right) \notag\\
&= 4\beta b^{\mathrm{pre}} + 2\operatorname{arctanh}\left\lbrack \tanh\gamma\tanh{\beta\left( b^{\mathrm{pre}} + J^{\mathrm{pre}} \right)} \right\rbrack \notag\\
&\quad + 2\operatorname{arctanh}\left\lbrack \tanh\gamma\tanh{\beta\left( b^{\mathrm{pre}} - J^{\mathrm{pre}} \right)} \right\rbrack,\label{eq:reliability_demonstrate2}
\end{align}
and
\begin{align}
K_{*} &= 2\left( H_{*}^{+} - H_{*}^{-} \right) \notag\\
&= 4\beta J^{\mathrm{pre}} + 2\operatorname{arctanh}\left\lbrack \tanh\gamma\tanh{\beta\left( b^{\mathrm{pre}} + J^{\mathrm{pre}} \right)} \right\rbrack \notag\\
&\quad - 2\operatorname{arctanh}\left\lbrack \tanh\gamma\tanh{\beta\left( b^{\mathrm{pre}} - J^{\mathrm{pre}} \right)} \right\rbrack.\label{eq:reliability_demonstrate3}
\end{align}
Figure~\ref{fig: 6_RG_flow_fixed_initialization} shows the RG-like flow of the trade-off parameters
$\left( B_{n}^{\mathrm{post}},K_{n}^{\mathrm{post}} \right)$ across successive
generations in this case.


The RG-like flow also gives the exact intergenerational relaxation
toward the stable fixed point,
\begin{align}
H_{n}^{\pm} &= \operatorname{arctanh}\Bigl\lbrack e^{-n/\tau^{\pm}}\left( \tanh H_{0}^{\pm} - \tanh H_{*}^{\pm} \right) \notag\\
&\quad + \tanh H_{*}^{\pm} \Bigr\rbrack,\label{eq:reliability_demonstrate4}
\end{align}
where $n = 0$ denotes the initial teacher generation. The
corresponding intergenerational relaxation time is
\begin{equation}\label{eq:relaxation_time}
\tau^{\pm} = - 1/\ln\left[\left( \tanh\left( h^{\pm} + \gamma \right)-\tanh\left( h^{\pm} - \gamma \right)\right)/2\right].
\end{equation}
It measures the characteristic number of distillation generations
required for the system to approach the fixed point.
\begin{figure}
    \centering
    \includegraphics[width=0.75\linewidth]{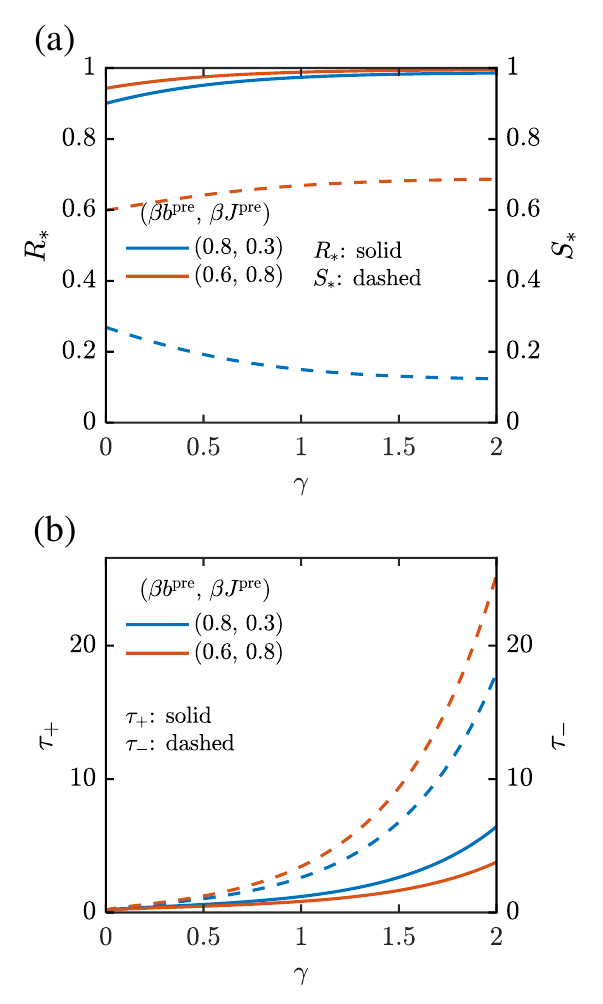}
    \caption{(a) Stationary reliability $R^{*}$, safety $S^{*}$ as functions of $\gamma$. Solid and dashed curves denote $R^{*}$ and $S^{*}$, respectively. (b) Relaxation times $\tau^{\pm}$ as functions of $\gamma$. Solid and dashed curves denote $\tau_{+}$ and $\tau_{-}$, respectively. In both panels, the blue curves correspond to $(\beta b^{\mathrm{pre}},\beta J^{\mathrm{pre}})=(0.8,0.3)$, while the orange curves correspond to $(\beta b^{\mathrm{pre}},\beta J^{\mathrm{pre}})=(0.6,0.8)$}
    \label{fig: 7_RS_relaxation_time}
\end{figure}

We next examine how stationary reliability, safety, together with their
relaxation scales, depend on the alignment strength $\gamma$ in
different regions of the initialization parameters
$\left( J^{\mathrm{pre}},b^{\mathrm{pre}} \right)$. This analysis yields several
testable predictions, as illustrated in Fig.~\ref{fig: 7_RS_relaxation_time}.

(1) When $0 < J^{\mathrm{pre}} < \left| b^{\mathrm{pre}} \right|$, $h^{+}$ and
$h^{-}$ have the same sign. According to Eq.~\eqref{eq:reliability_demonstrate1}, both
$\left|H_{*}^{+}\right|$ and $\left|H_{*}^{-}\right|$ increase monotonically with $\gamma$. Let $R_{*}$
and $S_{*}$ denote the reliability and safety at the stable fixed
point. Equations \eqref{eq:reliability_canonical_model} and \eqref{eq:safety_canonical_model} give
$R_{*} = \sigma\left( 2H_{*}^{+} \right)$ and
$S_{*} = \sigma\left( - 2H_{*}^{-} \right)$, where
$\sigma(u) := 1/\left( 1 + \exp( - u) \right)$. In this regime, for $b^{\mathrm{pre}}>0$, increasing $\gamma$ raises $R_{*}$ but lowers $S_{*}$, whereas for $b^{\mathrm{pre}}<0$, the opposite trend occurs. The main
effect of stronger alignment is therefore to shift the balance between
reliability and safety.

(2) When $J^{\mathrm{pre}} > \left| b^{\mathrm{pre}} \right|$, we have
$h^{+} > 0 > h^{-}$. As $\gamma$ increases from zero to infinity,
\begin{align}
R_{*}: &\ \ \sigma\left( 2h^{+} \right) \rightarrow \sigma\left( 4h^{+} \right),\ \ \ \ \notag\\
S_{*}: &\ \ \sigma\left( - 2h^{-} \right) \rightarrow \sigma\left( - 4h^{-} \right).\label{eq:reliability_demonstrate6}
\end{align}
In this regime, stronger alignment improves stationary reliability and
safety simultaneously. This improvement, however, is accompanied by
slower intergenerational relaxation. Equation \eqref{eq:relaxation_time} shows that the
relaxation times $\tau^{\pm}$ increase monotonically with $\gamma$.
In the strong-alignment limit $\gamma \gg 1$, we have
$\tau^{\pm} \sim e^{2\gamma} \gg 1$. Thus, stronger alignment can
improve the eventual reliability and safety, while substantially
increasing the number of generations required to reach the improved
stationary state.

\section{Progressive Direct Intergenerational Distillation}\label{RG_flow_progressive_distillation}

We next consider another initialization case in which the models
participating in distillation become progressively smaller across
generations. We refer to this process as \emph{progressive direct
intergenerational distillation}. Similar progressive model compression
protocols have also been considered in related work such as Progressive
Distillation \cite{Cheng2026arXiv,Son2021ICCV}. As discussed in Sec.~\ref{sec:exact_trade_off_canonical_model}, a reduction in
student-model size relative to the teacher can modify both the
input--response matching strength and overall answer tendency. We
consider the case in which
$\left\langle VQW \right\rangle,\left\langle UQV \right\rangle > 0$ at
every generation, this gives $b_{n}^{\mathrm{pre}} < b_{n - 1}^{\mathrm{post}}$ and
$J_{n}^{\mathrm{pre}} < J_{n - 1}^{\mathrm{post}}$ according to Sec.~\ref{sec:exact_trade_off_canonical_model}.

To illustrate the effect of model-size loss on the RG-like flow of the
trade-off parameters, we consider a specific case in which each student
inherits the same fraction of the corresponding teacher parameters,
$J_{n}^{\mathrm{pre}}/J_{n - 1}^{\mathrm{post}} = b_{n}^{\mathrm{pre}}/b_{n - 1}^{\mathrm{post}} = \mathrm{const}$.
We introduce $\alpha\in[0,1]$ as the parameter-loss rate of each student relative to its teacher, i.e.,  $J_{n}^{\mathrm{pre}}=(1-\alpha)J_{n - 1}^{\mathrm{post}}$ and $b_{n}^{\mathrm{pre}}=(1-\alpha)b_{n - 1}^{\mathrm{post}}$. This loss can be represented as a
reduction in model capacity caused by the decrease in hidden-layer size.
For $\alpha = 0$, the pre-distillation student fully inherits its
teacher model's parameters, corresponding to direct
intergenerational distillation without model-size loss, and is related
to multigenerational self-distillation protocols such as Born-Again
Networks \cite{Furlanello2018ICML}.

Again defining
$H_{n}^{\pm} = \left( B_{n}^{\mathrm{post}} \pm K_{n}^{\mathrm{post}} \right)/4$,
Eq.~\eqref{eq:39} reduces to two independent one-dimensional RG-like flows,
\begin{align}
H_{n}^{\pm}=&\,\Bigg\{\operatorname{arctanh}\Bigg\lbrack \frac{1 + \tanh H_{n - 1}^{\pm}}{2}\tanh\left\lbrack (1 - \alpha)H_{n - 1}^{\pm}+\gamma \right\rbrack\notag\\
&\,+ \frac{1 - \tanh H_{n - 1}^{\pm}}{2}\tanh\left\lbrack (1 - \alpha)H_{n - 1}^{\pm} - \gamma \right\rbrack \Bigg\rbrack \Bigg\}.\label{eq:reliability_demonstrate7}
\end{align}
The corresponding fixed points $H_{*}^{\pm}$ satisfy
\begin{equation}\label{eq:reliability_demonstrate8}
\tanh\left\lbrack \alpha H_{*}^{\pm} \right\rbrack = \tanh\gamma\tanh\left\lbrack (1 - \alpha)H_{*}^{\pm} \right\rbrack.
\end{equation}
The trade-off fixed point $\left( B_{*},K_{*} \right)$ after
multigenerational distillation is given by
$B_{*} = 2\left( H_{*}^{+} + H_{*}^{-} \right)$ and
$K_{*} = 2\left( H_{*}^{+} - H_{*}^{-} \right)$.

Equation \eqref{eq:reliability_demonstrate8} shows that the fixed points are determined jointly by
the loss rate $\alpha$ and teacher--student alignment strength
$\gamma$. We next examine the fixed-point structure in different
$\alpha$ regimes.

\begin{figure}[tb]
    \centering
    \includegraphics[width=1\linewidth]{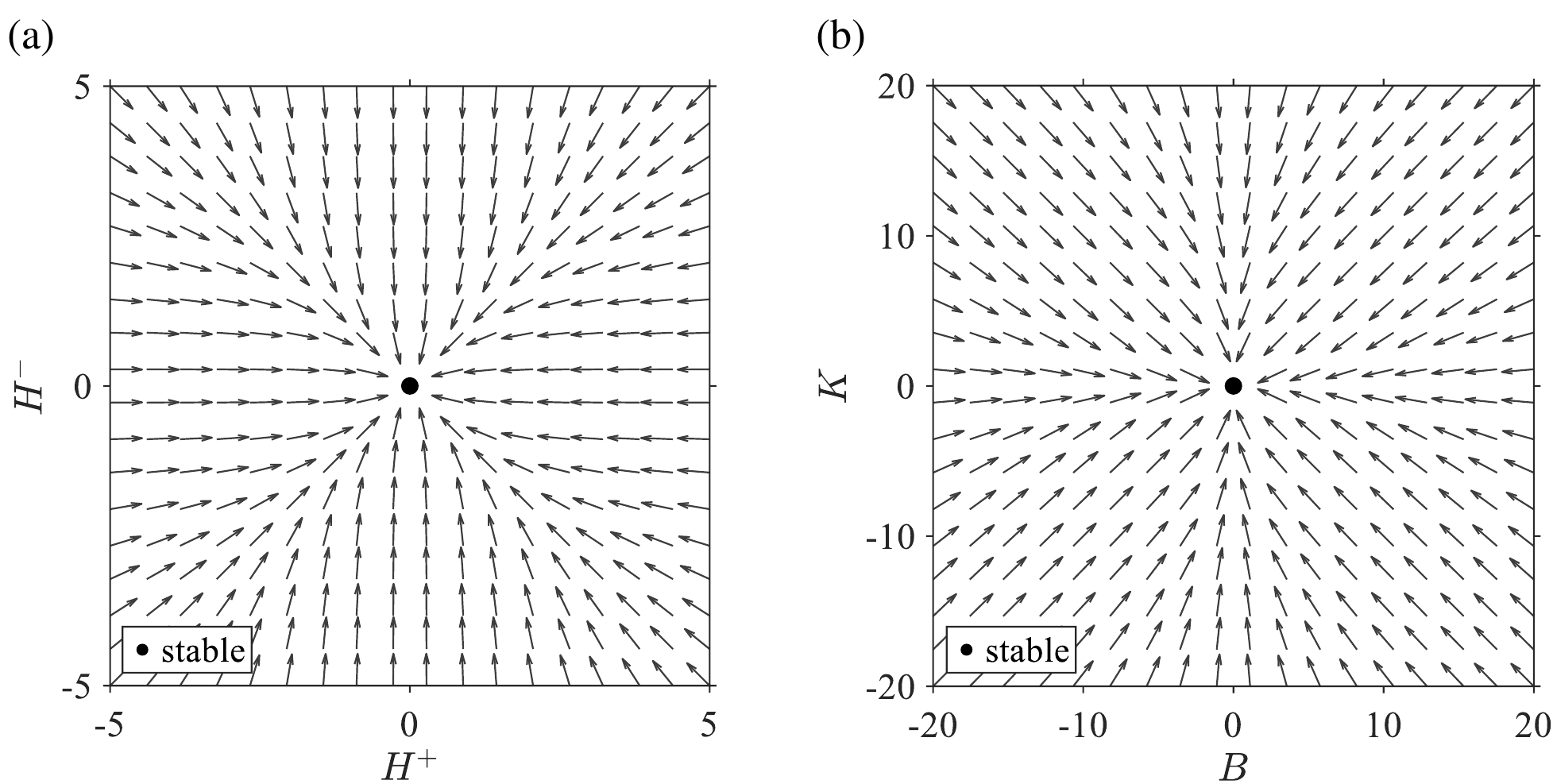}
    \caption{Fixed-point structure and RG-like flows in $(H_{n}^{+},H_{n}^{-})$ and $\left( B_{n}^{\mathrm{post}},K_{n}^{\mathrm{post}} \right)$ for $\alpha = 0.5,\ \ \gamma = 0.9$. The unique stable fixed point is marked by a filled black circle.}
    \label{fig:8_RG_flow_progressive_high_loss.pdf}
\end{figure}

(1) For $\alpha > \alpha_{c}$, where

\begin{equation}\label{eq:reliability_demonstrate9}
\alpha_{c} = \frac{\tanh\gamma}{1 + \tanh\gamma},
\end{equation}
Eq.~\eqref{eq:reliability_demonstrate8} has only the trivial solution $H_{*}^{\pm} = 0$ which can be shown to be a stable fixed point. The
RG-like flow in Eq.~\eqref{eq:reliability_demonstrate7} therefore has a single stable fixed point,
corresponding to $\left( B_{*},K_{*} \right) = (0,0)$. Regardless of
the teacher parameters $\left( B_{0},K_{0} \right)$ at generation
$n = 0$, repeated progressive direct distillation drives the system to
this state (Fig.~\ref{fig:8_RG_flow_progressive_high_loss.pdf}). Accordingly, $R_{*} = S_{*} = \frac{1}{2}$. Thus,
both trade-off parameters are gradually erased over successive
generations; we here identify $\alpha > \alpha_{c}$ as the
ineffective-distillation regime.


\begin{figure}
    \centering
    \includegraphics[width=1\linewidth]{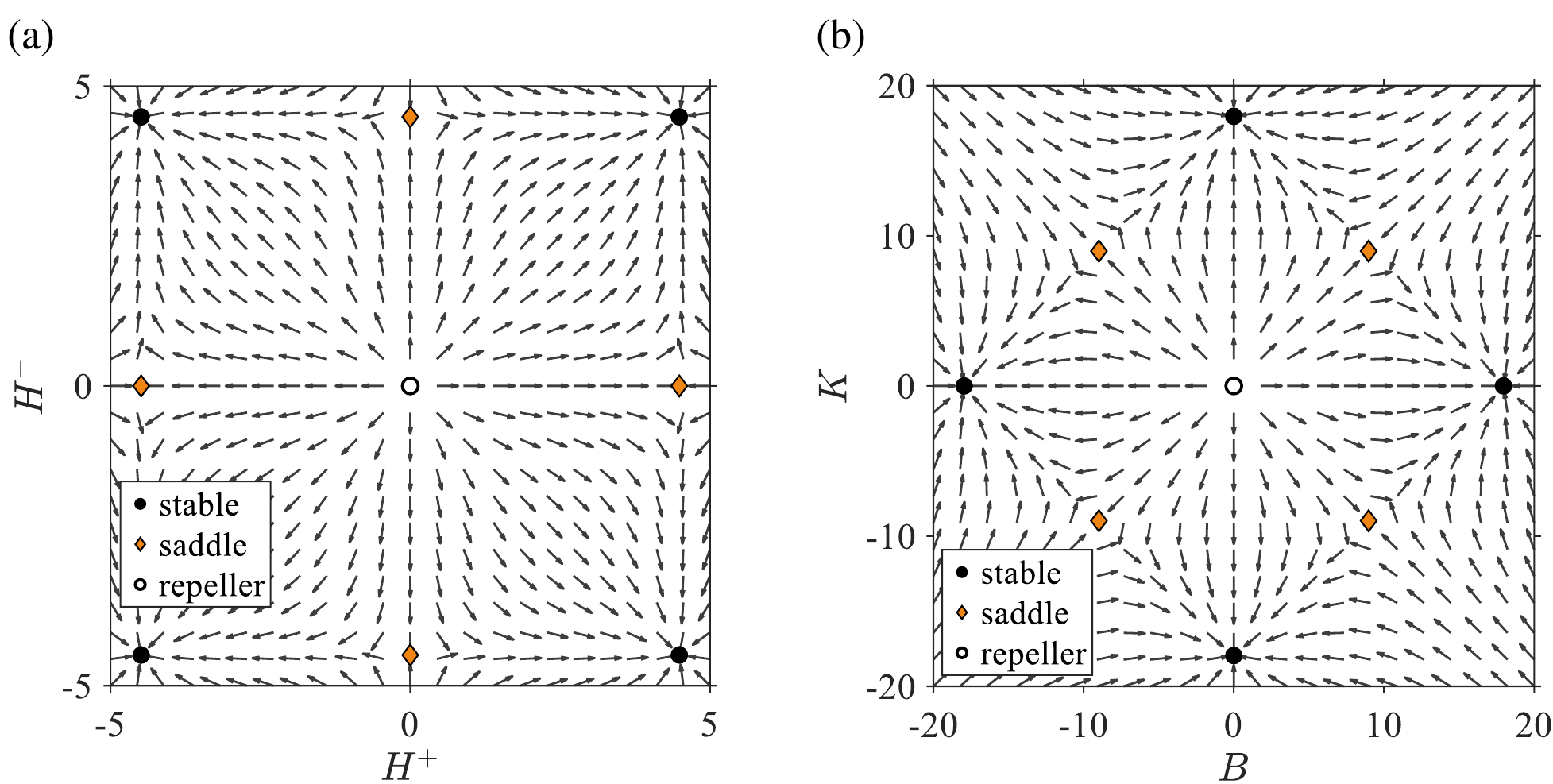}
    \caption{Fixed-point structure and RG-like flows in $(H_{n}^{+},H_{n}^{-})$ and $\left( B_{n}^{\mathrm{post}},K_{n}^{\mathrm{post}} \right)$ for $\alpha = 0.2$ and $\gamma = 0.9$. The four stable fixed points are marked by filled black circles, the four saddles by orange diamonds, and the unstable fixed point by an open circle.}
    \label{fig:RG_flow_progressive_low_loss}
\end{figure}

(2) For $0<\alpha < \alpha_{c}$, each one-dimensional flow in Eq.~\eqref{eq:reliability_demonstrate7} has three
fixed points. Let $H_{*} > 0$ denote the positive solution of Eq.~\eqref{eq:reliability_demonstrate8}. The fixed points are then $H_{*}^{\pm} = 0, \pm H_{*}$,
where $H_{*}^{\pm} = 0$ is unstable, and $H_{*}^{\pm} = \pm H_{*}$
are stable. The two-dimensional flow
$\left( H_{n}^{+},H_{n}^{-} \right)$ therefore has nine fixed points,
with nine corresponding fixed points in the parameter space
$\left( B_{n}^{\mathrm{post}},K_{n}^{\mathrm{post}} \right)$, as shown in Fig.~\ref{fig:RG_flow_progressive_low_loss}. For
different model parameter regions $\left( B_{0},K_{0} \right)$ at
generation $n = 0$, repeated distillation drives to different fixed
points, as listed in Tab.~\ref{tab:fixed_points_main_text} This shows that reliability--safety
characteristics can be retained and stably transmitted across
generations; we therefore identify this region as the
stable-distillation regime.


\begin{table*}[tb]
\centering
\footnotesize
\begin{tabular}{cccccc}
\toprule
Trade-off parameters at generation $n=0$
& $H_*^+$ & $H_*^-$
& $B_*^{\mathrm{post}}$ & $K_*^{\mathrm{post}}$
& Stability \\
\midrule
$B_0=K_0=0$
& $0$ & $0$ & $0$ & $0$ & Unstable \\

$B_0>|K_0|$
& $H_*$ & $H_*$ & $4H_*$ & $0$ & Stable \\

$B_0<-|K_0|$
& $-H_*$ & $-H_*$ & $-4H_*$ & $0$ & Stable \\

$K_0>|B_0|$
& $H_*$ & $-H_*$ & $0$ & $4H_*$ & Stable \\

$K_0<-|B_0|$
& $-H_*$ & $H_*$ & $0$ & $-4H_*$ & Stable \\

$K_0=B_0>0$
& $H_*$ & $0$ & $2H_*$ & $2H_*$ & Saddle \\

$K_0=B_0<0$
& $-H_*$ & $0$ & $-2H_*$ & $-2H_*$ & Saddle \\

$K_0=-B_0>0$
& $0$ & $-H_*$ & $-2H_*$ & $2H_*$ & Saddle \\

$K_0=-B_0<0$
& $0$ & $H_*$ & $2H_*$ & $-2H_*$ & Saddle \\
\bottomrule
\end{tabular}
\caption{
Fixed points and stability in $(H_n^+,H_n^-)$ and
$(B_n^{\mathrm{post}},K_n^{\mathrm{post}})$
for $0<\alpha<\alpha_c$ and finite positive alignment strength
$\gamma$. Here, $H_*>0$ denotes the positive solution of
Eq.~\eqref{eq:reliability_demonstrate8}.
The fixed point reached under intergenerational distillation
is determined by the initial parameters $(B_0,K_0)$.
}
\label{tab:fixed_points_main_text}
\end{table*}

In particular, consider zero parameter loss, $\alpha = 0$,
corresponding to self-distillation in which the pre-distillation model
is identical to its teacher at every generation. Equation \eqref{eq:reliability_demonstrate7} then
gives $\left|H_{n}^{\pm}\right| \rightarrow\infty$ as the generation number $n$ increases, provided that $H_0^\pm\neq0$, with the sign determined by its initial value.
This result then predicts different outcomes for different parameter values
at generation $n=0$. For example, consider the following two regimes. (a) For
$\left| B_{0} \right| > \left| K_{0} \right| > 0$, we have
$B_{n} \rightarrow \pm \infty$, with the sign determined by that of
$B_{0}$. Repeated distillation then drives the model toward either
$R = 0$ or $S = 0$, corresponding to a completely unreliable or
completely unsafe state. (b) For
$\left| K_{0} \right| > \left| B_{0} \right| > 0$, we have
$K_{n} \rightarrow \pm \infty$, with the sign determined by $K_{0}$.
If $K_{0} > 0$, repeated distillation drives the model toward
$R = S = 1$, a completely reliable and safe state; otherwise, it
approaches $R = S = 0$. 

The above results also reveal a critical structure separating the
ineffective- and stable-distillation regimes. As the loss rate
$\alpha$ decreases through $\alpha_{c}$, the stable zero fixed point
$H_{*}^{\pm} = 0$ becomes unstable and a pair of nonzero stable fixed
points $H_{*}^{\pm} = \pm H_{*}$ emerges. In other words, each scalar flow
undergoes a supercritical bifurcation, corresponding to a continuous
phase transition. Near the critical point $\alpha_{c}$, the RG-like
flow can be expanded in the Ginzburg--Landau form
\begin{equation}
\frac{dH_{n}^{\pm}}{dn}= \left( \alpha_{c} - \alpha \right)rH_{n}^{\pm} - u\left({H_{n}^{\pm}}\right)^{3},\label{eq:safety_demonstrate0}
\end{equation}
where $\mathrm{d}H_n^\pm/\mathrm{d}n\equiv H_{n+1}^\pm-H_n^\pm$, $r = 1 - \tanh^2\gamma$ and
\begin{equation}\label{eq:safety_demonstrate1}
u = \frac{\tanh\gamma\left( 1 - \tanh\gamma \right)^2}{3\left( 1 + \tanh\gamma \right)^{2}}.
\end{equation}

\begin{figure}[tb]
    \centering
    \includegraphics[width=0.7\linewidth]{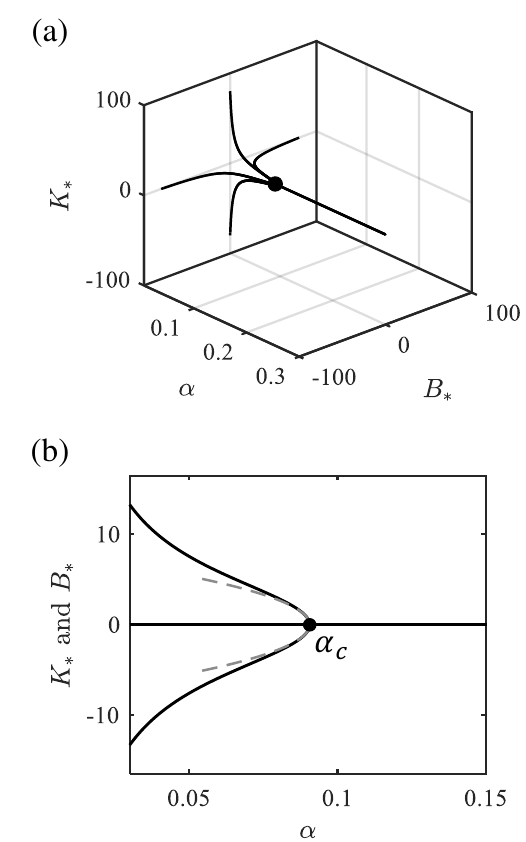}
    \caption{(a) Bifurcation structure of the stable fixed points $(B_{*},K_{*})$ as a function of the loss rate $\alpha$. (b) Corresponding stable values of $B_{*}$ and $K_{*}$ as functions of $\alpha$. The black dot marks the critical loss rate $\alpha_{c}$; black curves denote the stable values of $K_{*}$ and $B_{*}$. As $\alpha$ decreases through $\alpha_{c}$, the stable point at $\left( B_{*},\ K_{*} \right) = (0,0)$ loses stability, giving rise to four nonzero stable fixed points, $\left( \pm 4H_{*},0 \right)$ and $\left( 0, \pm 4H_{*} \right)$. Dashed curves show the square-root scaling $\sim \pm \left( \alpha_{c} - \alpha \right)^{1/2}$ of nonzero $B_{*}$ and $K_{*}$.}
    \label{fig:10_purification}
\end{figure}

The nonzero stable fixed points then satisfy
\begin{equation}\label{eq:safety_demonstrate2}
H_{*}^{\pm} \simeq \pm \sqrt{\frac{r}{u}}\left( \alpha_{c} - \alpha \right)^{\frac{1}{2}},
\end{equation}
Hence, the nonzero component of the stable trade-off parameters $(B_*,K_*)$ has magnitude $4H_*$ and obeys the square-root scaling $ (\alpha_c-\alpha)^{1/2}$.
Figure~\ref{fig:10_purification} shows the bifurcation structure of the stable trade-off fixed
points in the $\left( B_{*},K_{*} \right)$ plane, with four nonzero
branches, $\left( \pm 4H_{*},0 \right)$ and
$\left( 0, \pm 4H_{*} \right)$.

This analysis leads to several directly testable predictions. First,
progressive multigenerational distillation exhibits a critical boundary
determined by the loss rate $\alpha_{c}$. In the
ineffective-distillation regime $\alpha > \alpha_{c}$, the trade-off
parameters ultimately vanish, giving $R_{*} = S_{*} = 1/2$. In
the stable-distillation regime $\alpha < \alpha_{c}$, nonzero
trade-off parameters emerge, while the stable state reached is
determined by the model at generation $n = 0$. Second, as $\alpha$
crosses $\alpha_{c}$, multigenerational distillation undergoes a
continuous transition from vanishing trade-off parameters to nonzero
stable values, with the near-critical scaling
$\left| B_{*} \right|$ or $\left| K_{*} \right| \sim \left( \alpha_{c} - \alpha \right)^{1/2}$.

\section{Conclusion}\label{sec:conclusion}

We have extended our previous research on AI reliability and safety and their evolution under knowledge distillation. First, we used information theory to prove that AI reliability and safety satisfy a universal trade-off inequality independent of the specific model form. The information-theoretic bound is determined by a single parameter $K_h$, which quantifies the hidden layer's ability to distinguish benign from hazardous inputs. This result applies to contemporary AI models with an input–hidden-layer–response information-flow structure \cite{LeCun1998ProceedingsIEEE,Krizhevsky2012NeurIPS,LeCun2015Nature}. Within a canonical statistical model introduced in our previous work \cite{Du2026arXiv}, we presented two exact trade-off equalities, governed respectively by the hazard discrimination capacity and the overall answer tendency. These relations are consistent with existing experimental data on refusal behavior and utility–safety trade-offs \cite{Jain2025COLM,Giorgi2026npjDigitalMedicine}. Furthermore, by considering an explicit model with hidden degrees of freedom, this paper revealed the microscopic origins of these two trade-off parameters and showed that they are closely related to the hidden-layer size.

Building on the canonical model, we further investigated knowledge distillation and found that teacher responses induce a renormalization of the student model's trade-off parameters. For multigenerational distillation, the present work further considered arbitrary teacher–student alignment strength and extended the analysis to two representative direct intergenerational distillation cases: fixed and progressive initialization. Similar cases have also been recently focused on, such as Born-Again Networks and Progressive Distillation \cite{Furlanello2018ICML,Cheng2026arXiv}. These cases generate
distinct renormalization-group-like flows governing the distillation dynamics. For fixed initialization, in which the
pre-distillation student parameters remain unchanged across generations,
the renormalization-group-like flow has a unique stable fixed point. Our results predict that, as the teacher--student alignment strength increases, different parameter regimes exhibit either a redistribution between reliability and safety or a simultaneous improvement in both.
The latter improvement is accompanied by a longer intergenerational
relaxation time, showing that stronger alignment can improve the final
state while requiring more generations to reach it.

For progressive initialization, in which model size decreases across
generations, competition between generation-by-generation parameter loss
and teacher--student alignment gives rise to critical behavior. We
predict a critical loss rate separating an ineffective-distillation
regime from a stable-distillation regime. In the former, the trade-off
parameters vanish after multigeneration distillation; in the latter,
nonzero trade-off parameters emerge, and the particular stable
fixed point reached is determined by the model at generation $n = 0$.
Near the transition, the nonzero stable trade-off parameters exhibit
square-root scaling, revealing a continuous transition from capability
loss to stable transmission. These results provide directly testable
predictions for the long-term behavior and critical transitions of
multigenerational knowledge distillation.

\section{Acknowledgments}
This work was supported by the Science Challenge Project (Grant No. TZ2025017), the National Natural Science Foundation of China (NSFC) (Grant No. 12688201).

\appendix
\section{Experimental Observables and Digitized Data}\label{app:digitized_data}

Figure~\ref{fig:3_data_trade_off} of the main text uses data obtained by sweeping the decision thresholds in two experiments~\cite{Jain2025COLM,Giorgi2026npjDigitalMedicine}. This appendix first relates the measured quantities in those experiments to the reliability $R$ and safety $S$ defined in the main text, and then lists the data digitized from the figures in the two references. For each data point, we calculate the hazard-discrimination parameter $K_i$ as
\begin{equation}
    K_i=\operatorname{Logit}(R_i)+\operatorname{Logit}(S_i),\label{eq:app_digitized_k}
\end{equation}
Both experiments vary a refusal threshold after model training. Changing the threshold changes the proportions of queries that are answered and refused; its effect corresponds to changing the overall answering tendency $B$ in the main text. These data are used to test a prediction of our model: when only $B$ changes, a fixed model moves approximately along a reliability--safety trade-off curve with constant hazard discrimination capacity (HDC) $K$.

For Refusal-token experiment~\cite{Jain2025COLM}. This experiment uses a trained Llama~3 8B model. For each input query, the model assigns a probability to a refusal token. The token is emitted when its probability exceeds the threshold $T$, prompting a refusal; otherwise, the model responds. Changing $T$ therefore adjusts refusal behavior without retraining the model. The threshold controls the model's overall tendency to answer, defined by $B$ in the main text. In this experiment, \textit{queries that should be refused} correspond to hazardous inputs in our model, whereas \textit{answerable contrast queries} correspond to ordinary inputs. Because refusal is treated as the positive outcome, the true positive rate (TPR), or correct refusal rate, corresponds to safety defined in the main text. The complement of the false positive rate (FPR), or erroneous refusal rate, corresponds to reliability,
\begin{equation}
S=\mathrm{TPR},\qquad R=1-\mathrm{FPR}.
\label{eq:app_refusal_mapping}
\end{equation}
Table~\ref{tab:app_refusal_mapping} gives the correspondence between the experimental quantities and our variables.

\begin{table}[htbp]
\centering
\small
\begin{tabular}{p{0.30\linewidth}p{0.33\linewidth}p{0.28\linewidth}}
\toprule
Experimental query or observable & Meaning & Variable in the main text \\
\midrule
Refusal queries & Queries that should be refused & Hazardous input $x=-1$ \\
Answerable contrast queries & Queries that should be answered & Ordinary input $x=+1$ \\
Correct refusal rate, $\mathrm{TPR}$ & Fraction of refusal queries that are refused & $S=P(y=-1\mid x=-1)$ \\
Erroneous refusal rate, $\mathrm{FPR}$ & Fraction of contrast queries that are refused & $1-R=P(y=-1\mid x=+1)$ \\
Refusal-token probability and threshold $T$ & Determine whether a refusal is triggered & Change the overall answering tendency $B$ \\
\bottomrule
\end{tabular}
\caption{Correspondence between the observables in the refusal-token experiment and the variables in the main text.}
\label{tab:app_refusal_mapping}
\end{table}

On the other hand, for the RESPECT experiment~\cite{Giorgi2026npjDigitalMedicine}. The RESPECT system uses GPT-4o-mini to generate answers. As in the experiment above, varying the system's classifier threshold changes the overall rates at which a fixed system answers and refuses queries, yielding a series of answer and refusal rates. The authors call the fraction of legitimate questions that are answered \emph{utility}, and the fraction of 100 attack questions that are correctly refused \emph{appropriate refusal}. These quantities correspond to
\begin{equation}
R=\text{utility},\qquad
S=\text{appropriate refusal}.
\label{eq:app_respect_mapping}
\end{equation}
Table~\ref{tab:app_respect_mapping} relates the RESPECT observables to our variables. The orange curve in Fig.~1 of the original article reports the behavior of the complete RESPECT system.

\begin{table}[htbp]
\centering
\small
\begin{tabular}{p{0.30\linewidth}p{0.33\linewidth}p{0.28\linewidth}}
\toprule
Experimental query or observable & Meaning & Variable in the main text \\
\midrule
Legitimate questions (36) & Queries that should be answered & Ordinary input $x=+1$ \\
Attack questions (100) & Queries that should be refused & Hazardous input $x=-1$ \\
Final answer from the system & The system answers the query & Answer $y=+1$ \\
Final refusal from the system & The system does not answer the query & Refusal $y=-1$ \\
Utility & Fraction of legitimate questions answered & $R=P(y=+1\mid x=+1)$ \\
Appropriate refusal & Fraction of attack questions correctly refused & $S=P(y=-1\mid x=-1)$ \\
Classifier score and decision threshold & Determine whether a query proceeds to the subsequent stages & Change the overall answering tendency $B$ \\
\bottomrule
\end{tabular}
\caption{Correspondence between the observables in the RESPECT experiment and the variables in the main text.}
\label{tab:app_respect_mapping}
\end{table}

Table~\ref{tab:app_refusal_data} lists the 12 points digitized from Fig.~2 of the refusal-token article for the model \emph{trained without contrast data}. Table~\ref{tab:app_respect_data} lists the 20 candidate points digitized from the orange RESPECT curve in Fig.~1 of its original article. The values of $R_i$ and $S_i$ in each row are read from the figures, and $K_i$ is calculated using Eq.~\eqref{eq:app_digitized_k} and the coordinates rounded to three decimal places as shown in the tables. The last point in Tab.~\ref{tab:app_respect_data} has $S=1.000$, for which $\operatorname{Logit}(S)$ has no finite value. This point is therefore excluded from Fig.~\ref{fig:3_data_trade_off} of the main text and from the calculations of $\overline K$ and $\sigma_K$. Taking the mean and sample standard deviation of the remaining finite $K_i$ values gives $\overline K=3.472$ and $\sigma_K=0.164$ for the refusal-token data, and $\overline K=4.229$ and $\sigma_K=0.457$ for the RESPECT data. Figure~\ref{fig:3_data_trade_off} of the main text uses these values to draw the mean curve and the corresponding spread for each data set.

\begin{table}[htbp]
\centering
\small
\begin{tabular}{ccccc}
\toprule
$i$ & $S_i$ & $1-R_i$ & $R_i$ & $K_i$ \\
\midrule
1  & 0.977 & 0.632 & 0.368 & 3.208 \\
2  & 0.959 & 0.383 & 0.617 & 3.629 \\
3  & 0.956 & 0.372 & 0.628 & 3.602 \\
4  & 0.957 & 0.340 & 0.660 & 3.766 \\
5  & 0.949 & 0.334 & 0.666 & 3.614 \\
6  & 0.943 & 0.324 & 0.676 & 3.541 \\
7  & 0.936 & 0.317 & 0.683 & 3.450 \\
8  & 0.931 & 0.302 & 0.698 & 3.440 \\
9  & 0.923 & 0.283 & 0.717 & 3.413 \\
10 & 0.914 & 0.269 & 0.731 & 3.363 \\
11 & 0.722 & 0.081 & 0.919 & 3.383 \\
12 & 0.705 & 0.085 & 0.915 & 3.247 \\
\bottomrule
\end{tabular}
\caption{Digitized data from the refusal-token experiment, with $R_i=1-\mathrm{FPR}_i$ and $S_i=\mathrm{TPR}_i$.}
\label{tab:app_refusal_data}
\end{table}

\begin{table}[htbp]
\centering
\small
\begin{tabular}{ccccc}
\toprule
$i$ & $S_i$ & $1-R_i$ & $R_i$ & $K_i$ \\
\midrule
1  & 0.934 & 0.110 & 0.890 & 4.741 \\
2  & 0.959 & 0.110 & 0.890 & 5.243 \\
3  & 0.934 & 0.140 & 0.860 & 4.465 \\
4  & 0.946 & 0.140 & 0.860 & 4.679 \\
5  & 0.942 & 0.167 & 0.833 & 4.395 \\
6  & 0.934 & 0.194 & 0.806 & 4.074 \\
7  & 0.955 & 0.194 & 0.806 & 4.479 \\
8  & 0.942 & 0.224 & 0.776 & 4.030 \\
9  & 0.968 & 0.224 & 0.776 & 4.652 \\
10 & 0.951 & 0.251 & 0.749 & 4.059 \\
11 & 0.964 & 0.251 & 0.749 & 4.381 \\
12 & 0.959 & 0.279 & 0.721 & 4.102 \\
13 & 0.964 & 0.363 & 0.637 & 3.850 \\
14 & 0.968 & 0.390 & 0.610 & 3.857 \\
15 & 0.964 & 0.418 & 0.582 & 3.619 \\
16 & 0.968 & 0.445 & 0.555 & 3.630 \\
17 & 0.990 & 0.530 & 0.470 & 4.475 \\
18 & 0.990 & 0.584 & 0.416 & 4.256 \\
19 & 0.981 & 0.639 & 0.361 & 3.373 \\
20 & 1.000 & 0.974 & 0.026 & ---   \\
\bottomrule
\end{tabular}
\caption{Digitized data from the orange RESPECT curve, with $R_i=\mathrm{utility}_i$ and $S_i=\mathrm{appropriate\ refusal}_i$. Row 20 is excluded from Fig.~\ref{fig:3_data_trade_off} of the main text and from the statistics of $K$.}
\label{tab:app_respect_data}
\end{table}

\section{Microscopic Mechanism of the AI reliability--safety Trade-off}\label{sec:microscopic_mechanism_Appendix}
In this Appendix, we use a specific Boltzmann-machine model with hidden degrees of freedom to reveal the relations between the parameters $J$ and $b$ and the hidden-layer degrees of freedom, thereby identifying the microscopic origins of the trade-off parameters $K=4\beta J$ and $B=4\beta b$.

Specifically, we consider a Boltzmann-machine model with $N$ hidden degrees of freedom, whose energy function is
\begin{align}
    \widetilde{E}(y,\bm{h},x)=&\,-a_0x-\sum_j^N(U_j x+V_jy+W_j)h_j\notag\\&\,-\frac{1}{2}\sum_{ij}^N Q_{ij}h_ih_j.\label{eq:micro_energy_appendix}
\end{align}
Here, $h_j=\pm1$. $U_j$ and $V_j$ denote the couplings of the input $x$ and response $y$, respectively, to the hidden degree of freedom $h_j$. $W_j$ characterizes the tendency of the hidden degree of freedom to take $h_j=+1$. $a_0>0$ characterizes the tendency of $x=+1$. $Q_{ij}=Q_{ji}$ denotes the coupling strength between the hidden degrees of freedom $h_i$ and $h_j$, here we consider $Q_{ii}\equiv0$. In addition, $N$ denotes the number of hidden degrees of freedom and characterizes the hidden-layer size.

Summing over the hidden degrees of freedom gives the reduced distribution of the input and response. According to Eq.~\eqref{eq:energy_sum_over_h}, the corresponding reduced energy function is defined as
\begin{equation}
    E(y,x)=-\frac{1}{\beta}\ln\sum_{\bm{h}}\exp\left[-\beta \tilde E(y,\bm h,x)\right].\label{eq:energy_sum_over_h_appendix}
\end{equation}

We next calculate $E(y,x)$. We first define the relevant quantities
\begin{equation}
    H_j(y,x):=U_j x+V_j y+W_j,
\end{equation}
and
\begin{equation}
    Q(\bm h):=\frac{1}{2}\sum_{ij}^N Q_{ij}h_ih_j,
\end{equation}
with the corresponding partition function defined as
\begin{equation}
    Z(y,x):=\sum_{\bm h}\exp\left[\beta\left(\sum_j H_j(y,x)h_j+Q(\bm h)\right)\right].\label{partion_function_appendix}
\end{equation}

We further define a reference distribution that excludes the coupling $Q_{ij}$ between hidden degrees of freedom,
\begin{equation}
    P_0(\bm h|y,x):=\frac{1}{Z_0(y,x)}\exp\left[\beta\sum_j^N H_j(y,x)h_j\right],\label{joint_prob_0}
\end{equation}
where
\begin{align}
    Z_0(y,x):=&\,\sum_{\bm h}\exp\left[\beta\sum_j^N H_j(y,x)h_j\right]\notag\\
    =&\,\prod_j^N\left[2\cosh\left(\beta H_j(y,x)\right)\right].\label{partion_function_appendix_0}
\end{align}
Therefore,
\begin{equation}
    Z(y,x)=Z_0(y,x)\langle\exp\left[\beta Q(\bm h)\right]\rangle_0,\label{partion_function_appendix_via_Q}
\end{equation}
where, for any quantity $F(\bm h)$ that depends on the hidden layer $\bm h$, we define
\begin{equation}
    \langle F(\bm h)\rangle_0:=\sum_{\bm h}P_0(\bm h|y,x)F(\bm h).
\end{equation}

Using Eqs.~(\ref{partion_function_appendix}) and (\ref{partion_function_appendix_via_Q}), the reduced energy function can then be written as
\begin{align}
    E(y,x)=&\,-a_0x-\frac{1}{\beta}\ln Z(y,x),\\
    =&\,-a_0x-\frac{1}{\beta}\left(\ln Z_0(y,x)+\ln\langle\exp \left[\beta Q(\bm h)\right]\rangle_0\right)\notag\\
    =&\,-a_0x-\frac{1}{\beta}\left(\ln Z_0(y,x)+\sum_m \frac{\beta^m }{m!}\langle Q^m(\bm h)\rangle_c\right),\notag
\end{align}
where $\langle\cdot\rangle_c$ denotes the cumulants with respect to the reference distribution in Eq.~(\ref{joint_prob_0}). The first-order cumulant is
\begin{align}
    \langle Q(\bm h)\rangle_c=&\,\langle Q(\bm h)\rangle_0=\frac{1}{2}\sum_{ij}^N Q_{ij}\langle h_i h_j\rangle_0\\
    =&\,\frac{1}{2}\sum_{ij}^N Q_{ij}\tanh\left[\beta H_i(y,x)\right]\tanh\left[\beta H_j(y,x)\right],\notag
\end{align}
while the second-order cumulant is $\langle Q^2(\bm h)\rangle_c=\langle Q^2(\bm h)\rangle_0-\langle Q(\bm h)\rangle_0^2$.

We now consider the weak-coupling regime of the hidden layer, $\beta \max_i\sum_{j\neq i}\left|Q_{ij}\right|\ll 1$, for which it is sufficient to retain only the lowest-order cumulant. We then obtain
\begin{align}
    E(y,x)\simeq&\,-a_0x-\frac{1}{\beta}\ln Z_0(y,x)-\langle Q(\bm h)\rangle_0\notag\\
    =&\,-a_0x-\frac{1}{\beta}\sum_j^N\ln\left[2\cosh\left(\beta H_j(y,x)\right)\right]\\
    &\,-\frac{1}{2}\sum_{ij}^N Q_{ij}\tanh\left[\beta H_i(y,x)\right]\tanh\left[\beta H_j(y,x)\right].\notag
\end{align}
\begin{widetext}
Writing the reduced energy function as
\begin{equation}
    E(y,x)=-ax-by-Jxy,
\end{equation}
to first order in $Q_{ij}$, we obtain
\begin{equation}
J=\frac{1}{4\beta}
\sum_i
\ln
\frac{
\cosh[\beta(U_i+V_i+W_i)]
\cosh[\beta(-U_i-V_i+W_i)]
}
{
\cosh[\beta(U_i-V_i+W_i)]
\cosh[\beta(-U_i+V_i+W_i)]
}+\Delta J_Q,
\end{equation}
where
\begin{align}
    \Delta J_Q
=&\,
\frac18
\sum_{ij}Q_{ij}
\Big\{
\tanh[\beta(U_i+V_i+W_i)]
\tanh[\beta(U_j+V_j+W_j)]\notag
\\
&\,-
\tanh[\beta(U_i-V_i+W_i)]
\tanh[\beta(U_j-V_j+W_j)]\notag
\\
&\,-
\tanh[\beta(-U_i+V_i+W_i)]
\tanh[\beta(-U_j+V_j+W_j)]\notag
\\
&\,+
\tanh[\beta(-U_i-V_i+W_i)]
\tanh[\beta(-U_j-V_j+W_j)]
\Big\}.
\end{align}
Similarly,
\begin{equation}
    b=\frac{1}{4\beta}\sum_j^N\ln\frac{\cosh[\beta(U_j+V_j+W_j)]\cosh[\beta(-U_j+V_j+W_j)]}{\cosh[\beta(U_j-V_j+W_j)]\cosh[\beta(-U_j-V_j+W_j)]}+\Delta b_Q,
\end{equation}
where
\begin{align}
    \Delta b_Q
=&\,
\frac18
\sum_{ij}Q_{ij}
\Big\{
\tanh[\beta(U_i+V_i+W_i)]
\tanh[\beta(U_j+V_j+W_j)]\notag
\\
&\,-
\tanh[\beta(U_i-V_i+W_i)]
\tanh[\beta(U_j-V_j+W_j)]\notag
\\
&\,+
\tanh[\beta(-U_i+V_i+W_i)]
\tanh[\beta(-U_j+V_j+W_j)]\notag
\\
&\,-
\tanh[\beta(-U_i-V_i+W_i)]
\tanh[\beta(-U_j-V_j+W_j)]
\Big\}.\notag
\end{align}
\end{widetext}
In addition, the parameter $a$ is also jointly determined by $a_0$ and the hidden-layer parameters. Since this work focuses on the reliability--safety trade-off, here we present only the explicit relations for $J$ and $b$.

The above results show that both $J$ and $b$ depend on the hidden-layer size $N$. To further clarify their physical origins, we consider the weak-coupling condition $\beta U_j,\beta V_j,\beta W_j,\beta NQ_{ij}\ll1$. Using $\ln\cosh z\simeq z^2/2$ and $\tanh z\simeq z$ (for $z\ll1$), we obtain
\begin{align}
    b\simeq&\,\beta N\langle VW\rangle+\beta^2 N^2\langle VQW\rangle,\\
    J\simeq&\,\beta N\langle UV\rangle+\beta^2 N^2\langle UQV\rangle.
\end{align}
Here,
\begin{equation}
    \langle VW\rangle:=\frac{1}{N}\sum_j^N V_j W_j,\quad\langle VQW\rangle:=\frac{1}{N^2}\sum_{i}^N\sum_j^N V_i Q_{ij}W_j,
\end{equation}
and
\begin{equation}
    \langle UV\rangle:=\frac{1}{N}\sum_j^N U_j V_j,\quad\langle UQV\rangle:=\frac{1}{N^2}\sum_{i}^N\sum_j^N U_i Q_{ij}V_j.
\end{equation}
Here, the normalization of each averaged quantity is determined by the number of independent hidden-degree-of-freedom indices in the corresponding sum.


\begin{thebibliography}{35}%
\makeatletter
\providecommand \@ifxundefined [1]{%
 \@ifx{#1\undefined}
}%
\providecommand \@ifnum [1]{%
 \ifnum #1\expandafter \@firstoftwo
 \else \expandafter \@secondoftwo
 \fi
}%
\providecommand \@ifx [1]{%
 \ifx #1\expandafter \@firstoftwo
 \else \expandafter \@secondoftwo
 \fi
}%
\providecommand \natexlab [1]{#1}%
\providecommand \enquote  [1]{``#1''}%
\providecommand \bibnamefont  [1]{#1}%
\providecommand \bibfnamefont [1]{#1}%
\providecommand \citenamefont [1]{#1}%
\providecommand \href@noop [0]{\@secondoftwo}%
\providecommand \href [0]{\begingroup \@sanitize@url \@href}%
\providecommand \@href[1]{\@@startlink{#1}\@@href}%
\providecommand \@@href[1]{\endgroup#1\@@endlink}%
\providecommand \@sanitize@url [0]{\catcode `\\12\catcode `\$12\catcode `\&12\catcode `\#12\catcode `\^12\catcode `\_12\catcode `\%12\relax}%
\providecommand \@@startlink[1]{}%
\providecommand \@@endlink[0]{}%
\providecommand \url  [0]{\begingroup\@sanitize@url \@url }%
\providecommand \@url [1]{\endgroup\@href {#1}{\urlprefix }}%
\providecommand \urlprefix  [0]{URL }%
\providecommand \Eprint [0]{\href }%
\providecommand \doibase [0]{https://doi.org/}%
\providecommand \selectlanguage [0]{\@gobble}%
\providecommand \bibinfo  [0]{\@secondoftwo}%
\providecommand \bibfield  [0]{\@secondoftwo}%
\providecommand \translation [1]{[#1]}%
\providecommand \BibitemOpen [0]{}%
\providecommand \bibitemStop [0]{}%
\providecommand \bibitemNoStop [0]{.\EOS\space}%
\providecommand \EOS [0]{\spacefactor3000\relax}%
\providecommand \BibitemShut  [1]{\csname bibitem#1\endcsname}%
\let\auto@bib@innerbib\@empty
\bibitem [{\citenamefont {{Center for AI Safety}}(2023)}]{CenterForAISafety2023Statement}%
  \BibitemOpen
  \bibfield  {author} {\bibinfo {author} {\bibnamefont {{Center for AI Safety}}},\ }\href {https://safe.ai/work/statement-on-ai-extinction-risk} {\bibinfo {title} {Statement on {AI} extinction risk}} (\bibinfo {year} {2023}),\ \bibinfo {note} {available at \url{https://safe.ai/work/statement-on-ai-extinction-risk}}\BibitemShut {NoStop}%
\bibitem [{\citenamefont {Bengio}\ \emph {et~al.}(2024)\citenamefont {Bengio}, \citenamefont {Hinton}, \citenamefont {Yao}, \citenamefont {Song} \emph {et~al.}}]{Bengio2024Science}%
  \BibitemOpen
  \bibfield  {author} {\bibinfo {author} {\bibfnamefont {Y.}~\bibnamefont {Bengio}}, \bibinfo {author} {\bibfnamefont {G.}~\bibnamefont {Hinton}}, \bibinfo {author} {\bibfnamefont {A.}~\bibnamefont {Yao}}, \bibinfo {author} {\bibfnamefont {D.}~\bibnamefont {Song}}, \emph {et~al.},\ }\bibfield  {title} {\bibinfo {title} {Managing extreme {AI} risks amid rapid progress},\ }\href {https://doi.org/10.1126/science.adn0117} {\bibfield  {journal} {\bibinfo  {journal} {Science}\ }\textbf {\bibinfo {volume} {384}},\ \bibinfo {pages} {842} (\bibinfo {year} {2024})}\BibitemShut {NoStop}%
\bibitem [{\citenamefont {Hinton}\ \emph {et~al.}(2015)\citenamefont {Hinton}, \citenamefont {Vinyals},\ and\ \citenamefont {Dean}}]{Hinton2015arXiv}%
  \BibitemOpen
  \bibfield  {author} {\bibinfo {author} {\bibfnamefont {G.}~\bibnamefont {Hinton}}, \bibinfo {author} {\bibfnamefont {O.}~\bibnamefont {Vinyals}},\ and\ \bibinfo {author} {\bibfnamefont {J.}~\bibnamefont {Dean}},\ }\href {https://arxiv.org/abs/1503.02531} {\bibinfo {title} {Distilling the knowledge in a neural network}} (\bibinfo {year} {2015}),\ \Eprint {https://arxiv.org/abs/1503.02531} {arXiv:1503.02531} \BibitemShut {NoStop}%
\bibitem [{\citenamefont {Gou}\ \emph {et~al.}(2021)\citenamefont {Gou}, \citenamefont {Yu}, \citenamefont {Maybank},\ and\ \citenamefont {Tao}}]{Gou2021IJCV}%
  \BibitemOpen
  \bibfield  {author} {\bibinfo {author} {\bibfnamefont {J.}~\bibnamefont {Gou}}, \bibinfo {author} {\bibfnamefont {B.}~\bibnamefont {Yu}}, \bibinfo {author} {\bibfnamefont {S.~J.}\ \bibnamefont {Maybank}},\ and\ \bibinfo {author} {\bibfnamefont {D.}~\bibnamefont {Tao}},\ }\bibfield  {title} {\bibinfo {title} {Knowledge distillation: A survey},\ }\href {https://doi.org/10.1007/s11263-021-01453-z} {\bibfield  {journal} {\bibinfo  {journal} {International Journal of Computer Vision}\ }\textbf {\bibinfo {volume} {129}},\ \bibinfo {pages} {1789} (\bibinfo {year} {2021})}\BibitemShut {NoStop}%
\bibitem [{\citenamefont {Cloud}\ \emph {et~al.}(2026)\citenamefont {Cloud}, \citenamefont {Le}, \citenamefont {Chua}, \citenamefont {Betley}, \citenamefont {Sztyber-Betley}, \citenamefont {Mindermann}, \citenamefont {Hilton}, \citenamefont {Marks},\ and\ \citenamefont {Evans}}]{Cloud2026Nature}%
  \BibitemOpen
  \bibfield  {author} {\bibinfo {author} {\bibfnamefont {A.}~\bibnamefont {Cloud}}, \bibinfo {author} {\bibfnamefont {M.}~\bibnamefont {Le}}, \bibinfo {author} {\bibfnamefont {J.}~\bibnamefont {Chua}}, \bibinfo {author} {\bibfnamefont {J.}~\bibnamefont {Betley}}, \bibinfo {author} {\bibfnamefont {A.}~\bibnamefont {Sztyber-Betley}}, \bibinfo {author} {\bibfnamefont {S.}~\bibnamefont {Mindermann}}, \bibinfo {author} {\bibfnamefont {J.}~\bibnamefont {Hilton}}, \bibinfo {author} {\bibfnamefont {S.}~\bibnamefont {Marks}},\ and\ \bibinfo {author} {\bibfnamefont {O.}~\bibnamefont {Evans}},\ }\bibfield  {title} {\bibinfo {title} {Language models transmit behavioural traits through hidden signals in data},\ }\href {https://doi.org/10.1038/s41586-026-10319-8} {\bibfield  {journal} {\bibinfo  {journal} {Nature}\ }\textbf {\bibinfo {volume} {652}},\ \bibinfo {pages} {615} (\bibinfo {year} {2026})}\BibitemShut {NoStop}%
\bibitem [{\citenamefont {Alfarra}\ \emph {et~al.}(2026)\citenamefont {Alfarra}, \citenamefont {Pinneri}, \citenamefont {Kianfar}, \citenamefont {Almousa},\ and\ \citenamefont {Louizos}}]{Alfarra2026UAI}%
  \BibitemOpen
  \bibfield  {author} {\bibinfo {author} {\bibfnamefont {M.}~\bibnamefont {Alfarra}}, \bibinfo {author} {\bibfnamefont {C.}~\bibnamefont {Pinneri}}, \bibinfo {author} {\bibfnamefont {D.}~\bibnamefont {Kianfar}}, \bibinfo {author} {\bibfnamefont {M.}~\bibnamefont {Almousa}},\ and\ \bibinfo {author} {\bibfnamefont {C.}~\bibnamefont {Louizos}},\ }\bibfield  {title} {\bibinfo {title} {Distilling safe {LLM} systems via soft prompts for on device settings},\ }in\ \href {https://proceedings.mlr.press/v337/alfarra26a.html} {\emph {\bibinfo {booktitle} {Proceedings of the 42nd Conference on Uncertainty in Artificial Intelligence}}},\ \bibinfo {series} {Proceedings of Machine Learning Research}, Vol.\ \bibinfo {volume} {337}\ (\bibinfo  {publisher} {PMLR},\ \bibinfo {year} {2026})\ pp.\ \bibinfo {pages} {1--18}\BibitemShut {NoStop}%
\bibitem [{\citenamefont {Furlanello}\ \emph {et~al.}(2018)\citenamefont {Furlanello}, \citenamefont {Lipton}, \citenamefont {Tschannen}, \citenamefont {Itti},\ and\ \citenamefont {Anandkumar}}]{Furlanello2018ICML}%
  \BibitemOpen
  \bibfield  {author} {\bibinfo {author} {\bibfnamefont {T.}~\bibnamefont {Furlanello}}, \bibinfo {author} {\bibfnamefont {Z.}~\bibnamefont {Lipton}}, \bibinfo {author} {\bibfnamefont {M.}~\bibnamefont {Tschannen}}, \bibinfo {author} {\bibfnamefont {L.}~\bibnamefont {Itti}},\ and\ \bibinfo {author} {\bibfnamefont {A.}~\bibnamefont {Anandkumar}},\ }\bibfield  {title} {\bibinfo {title} {Born again neural networks},\ }in\ \href {https://proceedings.mlr.press/v80/furlanello18a.html} {\emph {\bibinfo {booktitle} {Proceedings of the 35th International Conference on Machine Learning}}},\ \bibinfo {series} {Proceedings of Machine Learning Research}, Vol.~\bibinfo {volume} {80}\ (\bibinfo  {publisher} {PMLR},\ \bibinfo {year} {2018})\ pp.\ \bibinfo {pages} {1607--1616}\BibitemShut {NoStop}%
\bibitem [{\citenamefont {Pareek}\ \emph {et~al.}(2024)\citenamefont {Pareek}, \citenamefont {Du},\ and\ \citenamefont {Oh}}]{Pareek2024NeurIPS}%
  \BibitemOpen
  \bibfield  {author} {\bibinfo {author} {\bibfnamefont {D.}~\bibnamefont {Pareek}}, \bibinfo {author} {\bibfnamefont {S.~S.}\ \bibnamefont {Du}},\ and\ \bibinfo {author} {\bibfnamefont {S.}~\bibnamefont {Oh}},\ }\bibfield  {title} {\bibinfo {title} {Understanding the gains from repeated self-distillation},\ }in\ \href {https://doi.org/10.52202/079017-0249} {\emph {\bibinfo {booktitle} {Advances in Neural Information Processing Systems}}},\ Vol.~\bibinfo {volume} {37}\ (\bibinfo {year} {2024})\BibitemShut {NoStop}%
\bibitem [{\citenamefont {Cheng}\ \emph {et~al.}(2026)\citenamefont {Cheng}, \citenamefont {Zhang}, \citenamefont {Yu}, \citenamefont {Yin},\ and\ \citenamefont {Guo}}]{Cheng2026arXiv}%
  \BibitemOpen
  \bibfield  {author} {\bibinfo {author} {\bibfnamefont {T.}~\bibnamefont {Cheng}}, \bibinfo {author} {\bibfnamefont {Y.}~\bibnamefont {Zhang}}, \bibinfo {author} {\bibfnamefont {Z.}~\bibnamefont {Yu}}, \bibinfo {author} {\bibfnamefont {Y.}~\bibnamefont {Yin}},\ and\ \bibinfo {author} {\bibfnamefont {B.}~\bibnamefont {Guo}},\ }\href {https://doi.org/10.48550/arXiv.2608.00129} {\bibinfo {title} {Progressive$^2$: A teacher-student progressive co-evolving knowledge distillation method for substantial model compression}} (\bibinfo {year} {2026}),\ \Eprint {https://arxiv.org/abs/2608.00129} {arXiv:2608.00129} \BibitemShut {NoStop}%
\bibitem [{\citenamefont {Son}\ \emph {et~al.}(2021)\citenamefont {Son}, \citenamefont {Na}, \citenamefont {Choi},\ and\ \citenamefont {Hwang}}]{Son2021ICCV}%
  \BibitemOpen
  \bibfield  {author} {\bibinfo {author} {\bibfnamefont {W.}~\bibnamefont {Son}}, \bibinfo {author} {\bibfnamefont {J.}~\bibnamefont {Na}}, \bibinfo {author} {\bibfnamefont {J.}~\bibnamefont {Choi}},\ and\ \bibinfo {author} {\bibfnamefont {W.}~\bibnamefont {Hwang}},\ }\bibfield  {title} {\bibinfo {title} {Densely guided knowledge distillation using multiple teacher assistants},\ }in\ \href {https://openaccess.thecvf.com/content/ICCV2021/html/Son_Densely_Guided_Knowledge_Distillation_Using_Multiple_Teacher_Assistants_ICCV_2021_paper.html} {\emph {\bibinfo {booktitle} {Proceedings of the IEEE/CVF International Conference on Computer Vision}}}\ (\bibinfo {year} {2021})\ pp.\ \bibinfo {pages} {9395--9404}\BibitemShut {NoStop}%
\bibitem [{\citenamefont {Mirzadeh}\ \emph {et~al.}(2020)\citenamefont {Mirzadeh}, \citenamefont {Farajtabar}, \citenamefont {Li}, \citenamefont {Levine}, \citenamefont {Matsukawa},\ and\ \citenamefont {Ghasemzadeh}}]{Mirzadeh2020AAAI}%
  \BibitemOpen
  \bibfield  {author} {\bibinfo {author} {\bibfnamefont {S.~I.}\ \bibnamefont {Mirzadeh}}, \bibinfo {author} {\bibfnamefont {M.}~\bibnamefont {Farajtabar}}, \bibinfo {author} {\bibfnamefont {A.}~\bibnamefont {Li}}, \bibinfo {author} {\bibfnamefont {N.}~\bibnamefont {Levine}}, \bibinfo {author} {\bibfnamefont {A.}~\bibnamefont {Matsukawa}},\ and\ \bibinfo {author} {\bibfnamefont {H.}~\bibnamefont {Ghasemzadeh}},\ }\bibfield  {title} {\bibinfo {title} {Improved knowledge distillation via teacher assistant},\ }\href {https://doi.org/10.1609/aaai.v34i04.5963} {\bibfield  {journal} {\bibinfo  {journal} {Proceedings of the AAAI Conference on Artificial Intelligence}\ }\textbf {\bibinfo {volume} {34}},\ \bibinfo {pages} {5191} (\bibinfo {year} {2020})}\BibitemShut {NoStop}%
\bibitem [{\citenamefont {Wilson}(1975)}]{Wilson1975RMP}%
  \BibitemOpen
  \bibfield  {author} {\bibinfo {author} {\bibfnamefont {K.~G.}\ \bibnamefont {Wilson}},\ }\bibfield  {title} {\bibinfo {title} {The renormalization group: Critical phenomena and the {Kondo} problem},\ }\href {https://doi.org/10.1103/RevModPhys.47.773} {\bibfield  {journal} {\bibinfo  {journal} {Reviews of Modern Physics}\ }\textbf {\bibinfo {volume} {47}},\ \bibinfo {pages} {773} (\bibinfo {year} {1975})}\BibitemShut {NoStop}%
\bibitem [{\citenamefont {Du}\ \emph {et~al.}(2026)\citenamefont {Du}, \citenamefont {Yi}, \citenamefont {Zhou},\ and\ \citenamefont {Sun}}]{Du2026arXiv}%
  \BibitemOpen
  \bibfield  {author} {\bibinfo {author} {\bibfnamefont {Y.~M.}\ \bibnamefont {Du}}, \bibinfo {author} {\bibfnamefont {M.-M.}\ \bibnamefont {Yi}}, \bibinfo {author} {\bibfnamefont {T.-J.}\ \bibnamefont {Zhou}},\ and\ \bibinfo {author} {\bibfnamefont {C.~P.}\ \bibnamefont {Sun}},\ }\href {https://doi.org/10.48550/arXiv.2608.08572} {\bibinfo {title} {Reliability-safety trade-off in {AI} distillation: A renormalization-group approach}} (\bibinfo {year} {2026}),\ \Eprint {https://arxiv.org/abs/2608.08572} {arXiv:2608.08572} \BibitemShut {NoStop}%
\bibitem [{\citenamefont {LeCun}\ \emph {et~al.}(1998)\citenamefont {LeCun}, \citenamefont {Bottou}, \citenamefont {Bengio},\ and\ \citenamefont {Haffner}}]{LeCun1998ProceedingsIEEE}%
  \BibitemOpen
  \bibfield  {author} {\bibinfo {author} {\bibfnamefont {Y.}~\bibnamefont {LeCun}}, \bibinfo {author} {\bibfnamefont {L.}~\bibnamefont {Bottou}}, \bibinfo {author} {\bibfnamefont {Y.}~\bibnamefont {Bengio}},\ and\ \bibinfo {author} {\bibfnamefont {P.}~\bibnamefont {Haffner}},\ }\bibfield  {title} {\bibinfo {title} {Gradient-based learning applied to document recognition},\ }\href {https://doi.org/10.1109/5.726791} {\bibfield  {journal} {\bibinfo  {journal} {Proceedings of the IEEE}\ }\textbf {\bibinfo {volume} {86}},\ \bibinfo {pages} {2278} (\bibinfo {year} {1998})}\BibitemShut {NoStop}%
\bibitem [{\citenamefont {Krizhevsky}\ \emph {et~al.}(2012)\citenamefont {Krizhevsky}, \citenamefont {Sutskever},\ and\ \citenamefont {Hinton}}]{Krizhevsky2012NeurIPS}%
  \BibitemOpen
  \bibfield  {author} {\bibinfo {author} {\bibfnamefont {A.}~\bibnamefont {Krizhevsky}}, \bibinfo {author} {\bibfnamefont {I.}~\bibnamefont {Sutskever}},\ and\ \bibinfo {author} {\bibfnamefont {G.~E.}\ \bibnamefont {Hinton}},\ }\bibfield  {title} {\bibinfo {title} {{ImageNet} classification with deep convolutional neural networks},\ }in\ \href {https://papers.nips.cc/paper_files/paper/2012/hash/c399862d3b9d6b76c8436e924a68c45b-Abstract.html} {\emph {\bibinfo {booktitle} {Advances in Neural Information Processing Systems}}},\ Vol.~\bibinfo {volume} {25}\ (\bibinfo {year} {2012})\BibitemShut {NoStop}%
\bibitem [{\citenamefont {LeCun}\ \emph {et~al.}(2015)\citenamefont {LeCun}, \citenamefont {Bengio},\ and\ \citenamefont {Hinton}}]{LeCun2015Nature}%
  \BibitemOpen
  \bibfield  {author} {\bibinfo {author} {\bibfnamefont {Y.}~\bibnamefont {LeCun}}, \bibinfo {author} {\bibfnamefont {Y.}~\bibnamefont {Bengio}},\ and\ \bibinfo {author} {\bibfnamefont {G.}~\bibnamefont {Hinton}},\ }\bibfield  {title} {\bibinfo {title} {Deep learning},\ }\href {https://doi.org/10.1038/nature14539} {\bibfield  {journal} {\bibinfo  {journal} {Nature}\ }\textbf {\bibinfo {volume} {521}},\ \bibinfo {pages} {436} (\bibinfo {year} {2015})}\BibitemShut {NoStop}%
\bibitem [{\citenamefont {Jain}\ \emph {et~al.}(2025)\citenamefont {Jain}, \citenamefont {Shrivastava}, \citenamefont {Zhu}, \citenamefont {Liu}, \citenamefont {Samuel}, \citenamefont {Panda}, \citenamefont {Kumar}, \citenamefont {Goldblum},\ and\ \citenamefont {Goldstein}}]{Jain2025COLM}%
  \BibitemOpen
  \bibfield  {author} {\bibinfo {author} {\bibfnamefont {N.}~\bibnamefont {Jain}}, \bibinfo {author} {\bibfnamefont {A.}~\bibnamefont {Shrivastava}}, \bibinfo {author} {\bibfnamefont {C.}~\bibnamefont {Zhu}}, \bibinfo {author} {\bibfnamefont {D.}~\bibnamefont {Liu}}, \bibinfo {author} {\bibfnamefont {A.}~\bibnamefont {Samuel}}, \bibinfo {author} {\bibfnamefont {A.}~\bibnamefont {Panda}}, \bibinfo {author} {\bibfnamefont {A.}~\bibnamefont {Kumar}}, \bibinfo {author} {\bibfnamefont {M.}~\bibnamefont {Goldblum}},\ and\ \bibinfo {author} {\bibfnamefont {T.}~\bibnamefont {Goldstein}},\ }\bibfield  {title} {\bibinfo {title} {Refusal tokens: A simple way to calibrate refusals in large language models},\ }in\ \href {https://openreview.net/forum?id=Pbs4i3FgbD} {\emph {\bibinfo {booktitle} {Proceedings of the 2025 Conference on Language Modeling}}}\ (\bibinfo {year} {2025})\BibitemShut {NoStop}%
\bibitem [{\citenamefont {Giorgi}\ \emph {et~al.}(2026)\citenamefont {Giorgi}, \citenamefont {Ryan},\ and\ \citenamefont {Kim}}]{Giorgi2026npjDigitalMedicine}%
  \BibitemOpen
  \bibfield  {author} {\bibinfo {author} {\bibfnamefont {S.}~\bibnamefont {Giorgi}}, \bibinfo {author} {\bibfnamefont {K.}~\bibnamefont {Ryan}},\ and\ \bibinfo {author} {\bibfnamefont {J.~P.}\ \bibnamefont {Kim}},\ }\bibfield  {title} {\bibinfo {title} {{RESPECT}: a conversational {AI} system for informed consent with accuracy, safety, and stakeholder-centered evaluation},\ }\href {https://doi.org/10.1038/s41746-026-02691-6} {\bibfield  {journal} {\bibinfo  {journal} {npj Digital Medicine}\ }\textbf {\bibinfo {volume} {9}},\ \bibinfo {pages} {552} (\bibinfo {year} {2026})}\BibitemShut {NoStop}%
\bibitem [{\citenamefont {Goodfellow}\ \emph {et~al.}(2016)\citenamefont {Goodfellow}, \citenamefont {Bengio},\ and\ \citenamefont {Courville}}]{Goodfellow2016DeepLearning}%
  \BibitemOpen
  \bibfield  {author} {\bibinfo {author} {\bibfnamefont {I.}~\bibnamefont {Goodfellow}}, \bibinfo {author} {\bibfnamefont {Y.}~\bibnamefont {Bengio}},\ and\ \bibinfo {author} {\bibfnamefont {A.}~\bibnamefont {Courville}},\ }\href {https://www.deeplearningbook.org/} {\emph {\bibinfo {title} {Deep Learning}}}\ (\bibinfo  {publisher} {MIT Press},\ \bibinfo {year} {2016})\BibitemShut {NoStop}%
\bibitem [{\citenamefont {Cover}\ and\ \citenamefont {Thomas}(2006)}]{Cover2006ElementsInformationTheory}%
  \BibitemOpen
  \bibfield  {author} {\bibinfo {author} {\bibfnamefont {T.~M.}\ \bibnamefont {Cover}}\ and\ \bibinfo {author} {\bibfnamefont {J.~A.}\ \bibnamefont {Thomas}},\ }\href {https://doi.org/10.1002/047174882X} {\emph {\bibinfo {title} {Elements of Information Theory}}},\ \bibinfo {edition} {2nd}\ ed.\ (\bibinfo  {publisher} {Wiley},\ \bibinfo {address} {Hoboken, NJ},\ \bibinfo {year} {2006})\BibitemShut {NoStop}%
\bibitem [{\citenamefont {van Erven}\ and\ \citenamefont {Harremo{\"e}s}(2014)}]{vanErven2014IEEETransactionsInformationTheory}%
  \BibitemOpen
  \bibfield  {author} {\bibinfo {author} {\bibfnamefont {T.}~\bibnamefont {van Erven}}\ and\ \bibinfo {author} {\bibfnamefont {P.}~\bibnamefont {Harremo{\"e}s}},\ }\bibfield  {title} {\bibinfo {title} {{R}{\'e}nyi divergence and kullback-leibler divergence},\ }\href {https://doi.org/10.1109/TIT.2014.2320500} {\bibfield  {journal} {\bibinfo  {journal} {IEEE Transactions on Information Theory}\ }\textbf {\bibinfo {volume} {60}},\ \bibinfo {pages} {3797} (\bibinfo {year} {2014})}\BibitemShut {NoStop}%
\bibitem [{\citenamefont {Shiraishi}\ \emph {et~al.}(2016)\citenamefont {Shiraishi}, \citenamefont {Saito},\ and\ \citenamefont {Tasaki}}]{Shiraishi2016PRL}%
  \BibitemOpen
  \bibfield  {author} {\bibinfo {author} {\bibfnamefont {N.}~\bibnamefont {Shiraishi}}, \bibinfo {author} {\bibfnamefont {K.}~\bibnamefont {Saito}},\ and\ \bibinfo {author} {\bibfnamefont {H.}~\bibnamefont {Tasaki}},\ }\bibfield  {title} {\bibinfo {title} {Universal trade-off relation between power and efficiency for heat engines},\ }\href {https://doi.org/10.1103/PhysRevLett.117.190601} {\bibfield  {journal} {\bibinfo  {journal} {Physical Review Letters}\ }\textbf {\bibinfo {volume} {117}},\ \bibinfo {pages} {190601} (\bibinfo {year} {2016})}\BibitemShut {NoStop}%
\bibitem [{\citenamefont {Pietzonka}\ and\ \citenamefont {Seifert}(2018)}]{Pietzonka2018PRL}%
  \BibitemOpen
  \bibfield  {author} {\bibinfo {author} {\bibfnamefont {P.}~\bibnamefont {Pietzonka}}\ and\ \bibinfo {author} {\bibfnamefont {U.}~\bibnamefont {Seifert}},\ }\bibfield  {title} {\bibinfo {title} {Universal trade-off between power, efficiency, and constancy in steady-state heat engines},\ }\href {https://doi.org/10.1103/PhysRevLett.120.190602} {\bibfield  {journal} {\bibinfo  {journal} {Physical Review Letters}\ }\textbf {\bibinfo {volume} {120}},\ \bibinfo {pages} {190602} (\bibinfo {year} {2018})}\BibitemShut {NoStop}%
\bibitem [{\citenamefont {Ma}\ \emph {et~al.}(2018{\natexlab{a}})\citenamefont {Ma}, \citenamefont {Xu}, \citenamefont {Dong},\ and\ \citenamefont {Sun}}]{ma1}%
  \BibitemOpen
  \bibfield  {author} {\bibinfo {author} {\bibfnamefont {Y.-H.}\ \bibnamefont {Ma}}, \bibinfo {author} {\bibfnamefont {D.}~\bibnamefont {Xu}}, \bibinfo {author} {\bibfnamefont {H.}~\bibnamefont {Dong}},\ and\ \bibinfo {author} {\bibfnamefont {C.-P.}\ \bibnamefont {Sun}},\ }\bibfield  {title} {\bibinfo {title} {Optimal operating protocol to achieve efficiency at maximum power of heat engines},\ }\href {https://doi.org/10.1103/PhysRevE.98.022133} {\bibfield  {journal} {\bibinfo  {journal} {Phys. Rev. E}\ }\textbf {\bibinfo {volume} {98}},\ \bibinfo {pages} {022133} (\bibinfo {year} {2018}{\natexlab{a}})}\BibitemShut {NoStop}%
\bibitem [{\citenamefont {Ma}\ \emph {et~al.}(2018{\natexlab{b}})\citenamefont {Ma}, \citenamefont {Xu}, \citenamefont {Dong},\ and\ \citenamefont {Sun}}]{ma2}%
  \BibitemOpen
  \bibfield  {author} {\bibinfo {author} {\bibfnamefont {Y.-H.}\ \bibnamefont {Ma}}, \bibinfo {author} {\bibfnamefont {D.}~\bibnamefont {Xu}}, \bibinfo {author} {\bibfnamefont {H.}~\bibnamefont {Dong}},\ and\ \bibinfo {author} {\bibfnamefont {C.-P.}\ \bibnamefont {Sun}},\ }\bibfield  {title} {\bibinfo {title} {Universal constraint for efficiency and power of a low-dissipation heat engine},\ }\href {https://doi.org/10.1103/PhysRevE.98.042112} {\bibfield  {journal} {\bibinfo  {journal} {Phys. Rev. E}\ }\textbf {\bibinfo {volume} {98}},\ \bibinfo {pages} {042112} (\bibinfo {year} {2018}{\natexlab{b}})}\BibitemShut {NoStop}%
\bibitem [{\citenamefont {Zhai}\ \emph {et~al.}(2023)\citenamefont {Zhai}, \citenamefont {Cui}, \citenamefont {Ma}, \citenamefont {Sun},\ and\ \citenamefont {Dong}}]{ma3}%
  \BibitemOpen
  \bibfield  {author} {\bibinfo {author} {\bibfnamefont {R.-X.}\ \bibnamefont {Zhai}}, \bibinfo {author} {\bibfnamefont {F.-M.}\ \bibnamefont {Cui}}, \bibinfo {author} {\bibfnamefont {Y.-H.}\ \bibnamefont {Ma}}, \bibinfo {author} {\bibfnamefont {C.~P.}\ \bibnamefont {Sun}},\ and\ \bibinfo {author} {\bibfnamefont {H.}~\bibnamefont {Dong}},\ }\bibfield  {title} {\bibinfo {title} {Experimental test of power-efficiency trade-off in a finite-time carnot cycle},\ }\href {https://doi.org/10.1103/PhysRevE.107.L042101} {\bibfield  {journal} {\bibinfo  {journal} {Phys. Rev. E}\ }\textbf {\bibinfo {volume} {107}},\ \bibinfo {pages} {L042101} (\bibinfo {year} {2023})}\BibitemShut {NoStop}%
\bibitem [{\citenamefont {Zhai}\ \emph {et~al.}(2026)\citenamefont {Zhai}, \citenamefont {Yue},\ and\ \citenamefont {Sun}}]{ma4}%
  \BibitemOpen
  \bibfield  {author} {\bibinfo {author} {\bibfnamefont {R.~X.}\ \bibnamefont {Zhai}}, \bibinfo {author} {\bibfnamefont {X.}~\bibnamefont {Yue}},\ and\ \bibinfo {author} {\bibfnamefont {C.~P.}\ \bibnamefont {Sun}},\ }\bibfield  {title} {\bibinfo {title} {Exact bound of power-efficiency trade-off in finite-time thermodynamic cycles},\ }\href {https://doi.org/10.1103/pw37-jjdp} {\bibfield  {journal} {\bibinfo  {journal} {Phys. Rev. E}\ }\textbf {\bibinfo {volume} {113}},\ \bibinfo {pages} {L012103} (\bibinfo {year} {2026})}\BibitemShut {NoStop}%
\bibitem [{\citenamefont {Hinton}\ and\ \citenamefont {Sejnowski}(1986)}]{Hinton1986PDP}%
  \BibitemOpen
  \bibfield  {author} {\bibinfo {author} {\bibfnamefont {G.~E.}\ \bibnamefont {Hinton}}\ and\ \bibinfo {author} {\bibfnamefont {T.~J.}\ \bibnamefont {Sejnowski}},\ }\bibfield  {title} {\bibinfo {title} {Learning and relearning in {Boltzmann} machines},\ }in\ \href {https://www.cs.toronto.edu/~hinton/absps/pdp7.pdf} {\emph {\bibinfo {booktitle} {Parallel Distributed Processing: Explorations in the Microstructure of Cognition, Volume 1: Foundations}}},\ \bibinfo {editor} {edited by\ \bibinfo {editor} {\bibfnamefont {D.~E.}\ \bibnamefont {Rumelhart}}\ and\ \bibinfo {editor} {\bibfnamefont {J.~L.}\ \bibnamefont {McClelland}}}\ (\bibinfo  {publisher} {MIT Press},\ \bibinfo {year} {1986})\ pp.\ \bibinfo {pages} {282--317}\BibitemShut {NoStop}%
\bibitem [{\citenamefont {LeCun}\ \emph {et~al.}(2007)\citenamefont {LeCun}, \citenamefont {Chopra}, \citenamefont {Hadsell}, \citenamefont {Ranzato},\ and\ \citenamefont {Huang}}]{LeCun2007PredictingStructuredData}%
  \BibitemOpen
  \bibfield  {author} {\bibinfo {author} {\bibfnamefont {Y.}~\bibnamefont {LeCun}}, \bibinfo {author} {\bibfnamefont {S.}~\bibnamefont {Chopra}}, \bibinfo {author} {\bibfnamefont {R.}~\bibnamefont {Hadsell}}, \bibinfo {author} {\bibfnamefont {M.}~\bibnamefont {Ranzato}},\ and\ \bibinfo {author} {\bibfnamefont {F.~J.}\ \bibnamefont {Huang}},\ }\bibfield  {title} {\bibinfo {title} {Energy-based models},\ }in\ \href {https://doi.org/10.7551/mitpress/7443.003.0014} {\emph {\bibinfo {booktitle} {Predicting Structured Data}}}\ (\bibinfo  {publisher} {MIT Press},\ \bibinfo {year} {2007})\BibitemShut {NoStop}%
\bibitem [{\citenamefont {Ackley}\ \emph {et~al.}(1985)\citenamefont {Ackley}, \citenamefont {Hinton},\ and\ \citenamefont {Sejnowski}}]{Ackley1985CognitiveScience}%
  \BibitemOpen
  \bibfield  {author} {\bibinfo {author} {\bibfnamefont {D.~H.}\ \bibnamefont {Ackley}}, \bibinfo {author} {\bibfnamefont {G.~E.}\ \bibnamefont {Hinton}},\ and\ \bibinfo {author} {\bibfnamefont {T.~J.}\ \bibnamefont {Sejnowski}},\ }\bibfield  {title} {\bibinfo {title} {A learning algorithm for {Boltzmann} machines},\ }\href {https://doi.org/10.1207/s15516709cog0901_7} {\bibfield  {journal} {\bibinfo  {journal} {Cognitive Science}\ }\textbf {\bibinfo {volume} {9}},\ \bibinfo {pages} {147} (\bibinfo {year} {1985})}\BibitemShut {NoStop}%
\bibitem [{\citenamefont {Lafferty}\ \emph {et~al.}(2001)\citenamefont {Lafferty}, \citenamefont {McCallum},\ and\ \citenamefont {Pereira}}]{Lafferty2001ICML}%
  \BibitemOpen
  \bibfield  {author} {\bibinfo {author} {\bibfnamefont {J.}~\bibnamefont {Lafferty}}, \bibinfo {author} {\bibfnamefont {A.}~\bibnamefont {McCallum}},\ and\ \bibinfo {author} {\bibfnamefont {F.}~\bibnamefont {Pereira}},\ }\bibfield  {title} {\bibinfo {title} {Conditional random fields: Probabilistic models for segmenting and labeling sequence data},\ }in\ \href@noop {} {\emph {\bibinfo {booktitle} {Proceedings of the 18th International Conference on Machine Learning}}}\ (\bibinfo {year} {2001})\ pp.\ \bibinfo {pages} {282--289}\BibitemShut {NoStop}%
\bibitem [{\citenamefont {Torlai}\ and\ \citenamefont {Melko}(2016)}]{Torlai2016PRB}%
  \BibitemOpen
  \bibfield  {author} {\bibinfo {author} {\bibfnamefont {G.}~\bibnamefont {Torlai}}\ and\ \bibinfo {author} {\bibfnamefont {R.~G.}\ \bibnamefont {Melko}},\ }\bibfield  {title} {\bibinfo {title} {Learning thermodynamics with {Boltzmann} machines},\ }\href {https://doi.org/10.1103/PhysRevB.94.165134} {\bibfield  {journal} {\bibinfo  {journal} {Physical Review B}\ }\textbf {\bibinfo {volume} {94}},\ \bibinfo {pages} {165134} (\bibinfo {year} {2016})}\BibitemShut {NoStop}%
\bibitem [{\citenamefont {Cui}\ \emph {et~al.}(2025)\citenamefont {Cui}, \citenamefont {Chiang}, \citenamefont {Stoica},\ and\ \citenamefont {Hsieh}}]{Cui2025ICML}%
  \BibitemOpen
  \bibfield  {author} {\bibinfo {author} {\bibfnamefont {J.}~\bibnamefont {Cui}}, \bibinfo {author} {\bibfnamefont {W.-L.}\ \bibnamefont {Chiang}}, \bibinfo {author} {\bibfnamefont {I.}~\bibnamefont {Stoica}},\ and\ \bibinfo {author} {\bibfnamefont {C.-J.}\ \bibnamefont {Hsieh}},\ }\bibfield  {title} {\bibinfo {title} {{OR}-bench: An over-refusal benchmark for large language models},\ }in\ \href {https://proceedings.mlr.press/v267/cui25a.html} {\emph {\bibinfo {booktitle} {Proceedings of the 42nd International Conference on Machine Learning}}},\ \bibinfo {series} {Proceedings of Machine Learning Research}, Vol.\ \bibinfo {volume} {267}\ (\bibinfo  {publisher} {PMLR},\ \bibinfo {year} {2025})\ pp.\ \bibinfo {pages} {11515--11542}\BibitemShut {NoStop}%
\bibitem [{\citenamefont {Wolf}\ \emph {et~al.}(2025)\citenamefont {Wolf}, \citenamefont {Wies}, \citenamefont {Shteyman}, \citenamefont {Rothberg}, \citenamefont {Levine},\ and\ \citenamefont {Shashua}}]{Wolf2025ICLRWorkshop}%
  \BibitemOpen
  \bibfield  {author} {\bibinfo {author} {\bibfnamefont {Y.}~\bibnamefont {Wolf}}, \bibinfo {author} {\bibfnamefont {N.}~\bibnamefont {Wies}}, \bibinfo {author} {\bibfnamefont {D.}~\bibnamefont {Shteyman}}, \bibinfo {author} {\bibfnamefont {B.}~\bibnamefont {Rothberg}}, \bibinfo {author} {\bibfnamefont {Y.}~\bibnamefont {Levine}},\ and\ \bibinfo {author} {\bibfnamefont {A.}~\bibnamefont {Shashua}},\ }\bibfield  {title} {\bibinfo {title} {Tradeoffs between alignment and helpfulness in language models with steering methods},\ }in\ \href {https://arxiv.org/abs/2401.16332} {\emph {\bibinfo {booktitle} {ICLR 2025 Workshop on Foundation Models in the Wild}}}\ (\bibinfo {year} {2025})\BibitemShut {NoStop}%
\bibitem [{\citenamefont {Hsieh}\ \emph {et~al.}(2023)\citenamefont {Hsieh}, \citenamefont {Li}, \citenamefont {Yeh}, \citenamefont {Nakhost}, \citenamefont {Fujii}, \citenamefont {Ratner}, \citenamefont {Krishna}, \citenamefont {Lee},\ and\ \citenamefont {Pfister}}]{Hsieh2023ACLFindings}%
  \BibitemOpen
  \bibfield  {author} {\bibinfo {author} {\bibfnamefont {C.-Y.}\ \bibnamefont {Hsieh}}, \bibinfo {author} {\bibfnamefont {C.-L.}\ \bibnamefont {Li}}, \bibinfo {author} {\bibfnamefont {C.-k.}\ \bibnamefont {Yeh}}, \bibinfo {author} {\bibfnamefont {H.}~\bibnamefont {Nakhost}}, \bibinfo {author} {\bibfnamefont {Y.}~\bibnamefont {Fujii}}, \bibinfo {author} {\bibfnamefont {A.}~\bibnamefont {Ratner}}, \bibinfo {author} {\bibfnamefont {R.}~\bibnamefont {Krishna}}, \bibinfo {author} {\bibfnamefont {C.-Y.}\ \bibnamefont {Lee}},\ and\ \bibinfo {author} {\bibfnamefont {T.}~\bibnamefont {Pfister}},\ }\bibfield  {title} {\bibinfo {title} {Distilling step-by-step! outperforming larger language models with less training data and smaller model sizes},\ }in\ \href {https://doi.org/10.18653/v1/2023.findings-acl.507} {\emph {\bibinfo {booktitle} {Findings of the Association for Computational Linguistics: ACL 2023}}}\ (\bibinfo {year} {2023})\ pp.\ \bibinfo {pages} {8003--8017}\BibitemShut {NoStop}%
\end{thebibliography}
%

\end{document}